\documentclass[fleqn,usenatbib]{mnras}

\usepackage[T1]{fontenc}

\DeclareRobustCommand{\VAN}[3]{#2}
\let\VANthebibliography\thebibliography
\def\thebibliography{\DeclareRobustCommand{\VAN}[3]{##3}\VANthebibliography}

\usepackage{graphicx}	
\usepackage{amsmath}	
\usepackage{amssymb}	

\title[The highly obscured nucleus in NGC 1448]{The highly obscured Seyfert 2 nucleus in NGC 1448 observed with MUSE}

\author[R. B. Menezes et al.]{
R. B. Menezes$^{1}$\thanks{E-mail: roberto.menezes@maua.br},
Patr\'icia da Silva$^{2}$\thanks{E-mail: p.silva2201@gmail.com}
and J. E. Steiner$^{2}$\thanks{Deceased}
\\
$^{1}$Instituto Mau\'a de Tecnologia, Pra\c{c}a Mau\'a 1, 09580-900, S\~ao Caetano do Sul, SP, Brazil\\
$^{2}$Instituto de Astronomia, Geof\'isica e Ci\^encias Atmosf\'ericas, Departamento de Astronomia, Universidade de S\~ao Paulo, 05508-090, SP, Brazil
}

\date{Accepted 2021 February 12. Received 2021 February 11; in original form 2021 February 2}

\pubyear{2021}

\begin{document}
\label{firstpage}
\pagerange{\pageref{firstpage}--\pageref{lastpage}}
\maketitle

\begin{abstract}

We present the analysis of an optical data cube of the central region of NGC 1448, obtained with the Multi Unit Spectroscopic Explorer (MUSE). Chandra X-ray data indicate that the active galactic nucleus (AGN) is not located at the apparent stellar nucleus of the galaxy, but at a projected distance of $1.75$ $\pm$ $0.22$ arcsec ($139 \pm 17$ pc). This is probably caused by the high interstellar extinction in the surroundings of the AGN, which corresponds to the true nucleus of the galaxy, as also proposed by previous studies. The morphology and classification of the optical line-emitting regions indicate two ionization cones, around an axis with a position angle of $PA_{cones} = -50\degr \pm 7\degr$, with emission-line spectra characteristic of Seyfert galaxies. The stellar and gas kinematics are consistent with a stellar and gas rotating disc around the nucleus, with a velocity amplitude of 125 km s$^{-1}$. Two probable outflows from the AGN were detected along the region of the two ionization cones. The AGN position does not coincide with the brightest line-emitting region at the centre of NGC 1448. That may be a consequence of the high obscuration from the AGN towards the observer (the AGN is actually Compton thick), mostly caused by a nearly edge-on torus. An additional hypothesis is that the AGN reduced its luminosity, during the last $440$ yr, to nearly half of the value in the past. In this case, the brightest line-emitting region corresponds to a ``light echo'' or a ``fossil'' of the AGN in the past. 

\end{abstract}




\section{Introduction}

Active galactic nuclei (AGNs) have a major relevance in extragalactic astrophysics. It is well accepted that the energy emitted by AGNs comes from the accretion of matter onto central supermassive black holes (SMBHs; e.g. \citealt{net13}). This feeding mechanism can also lead to feedback from the AGN, which can have a significant impact, for example, on the growth of the SMBH and on the star formation (e.g. \citealt{cro06}). The existence of correlations of parameters associated with the central SMBHs and with their host galaxies suggests a coevolution. An example of such correlations is the $M-\sigma$ relation \citep{fer00,geb00,gul09}, between the masses of the SMBHs ($M$) and the stellar velocity dispersions ($\sigma$) of the bulges of the corresponding galaxies. The feeding and feedback processes are probably essential to understand the coevolution of SMBHs and the host galaxies. In this context, having a complete census of the presence of AGNs in galaxies of different morphological types is very important for studies of the formation and evolution of galaxies.

Highly obscuring material along the line of sight can cause some difficulties in the detection of AGNs. Compton-thick AGNs, which are defined as AGNs obscured by a column density of $N_H \geq 1/\sigma_T = 1.5 \times 10^{24} cm^{-2}$ ($\sigma_T$ is the Thomson scattering cross-section), are the most challenging ones. In this case, the extinction can be significant even at high energies ($\geq 10$ keV). It is usually assumed that the obscuring material in a Compton-thick AGN is located in a torus around the broad-line region (BLR), according to the unified model \citep{ant93,urr95}. On the other hand, according to recent studies, the actual scenario around AGNs may be more complex than that. \citet{hoe19}, for example, based on observational properties of AGNs in the infrared (IR) and submm wavebands, proposed the existence of a gas flow from scales of $\sim$100 pc to the nuclear region via a disc, which, due to radiation pressure, puffs up close to the AGN. The radiation pressure acting on the disc generates a dusty wind. Therefore, according to this scenario, the angle-dependent obscuration predicted by the unified model results from the combination of the disc, wind, and also the wind launching region. Since the discussion of this complex scenario is beyond the scope of this work, we will be referring, throughout the paper, to the structure causing the obscuration in the surroundings of the AGN as the torus.  

Even in non-Compton-thick AGNs, the presence of obscuring material (e.g. dust lanes) along the line of sight, at larger distances from the nucleus than the torus, can also compromise the detection of AGNs, especially in the optical and ultraviolet (UV). In certain cases, not only the detection of the AGN but also the determination of the precise location of the galaxy's nucleus can be affected by obscuring material. \citet{pri14}, for example, analysed a sample of type 2 AGNs using optical Hubble Space Telescope (HST) images and high spatial resolution adaptive optics IR images, taken with the Very Large Telescope, which are less affected by dust extinction, and verified that, in most cases, the IR nucleus is displaced by tens of pc from the optical nucleus. The authors concluded that the IR nuclei were behind dust lanes and that the apparent displacement between the optical and the IR nucleus of each galaxy was a consequence of interstellar extinction caused by the material along the line of sight of the IR nucleus, which represents the true nucleus of the object. This difficulty for detecting obscured galactic nuclei in certain spectral bands may result in the determination of wrong parameters, such as the luminosity. For some objects, the detection of the kinematic centre (by the analysis of the stellar or gas kinematics) may allow the identification of the true nucleus. NGC 253 is a good example of a galaxy whose analysis of the central region (including the identification of the true nucleus) was considerably affected by dust extinction \citep{mul10}.

NGC 1448 is a late-type galaxy, classified as SAcd \citep{rc3}. The distance of this object that will be considered throughout this paper, corresponding to the mean of the redshift-independent distances provided by the NASA Extragalactic Database (NED), is 16.4 Mpc. However, it is worth mentioning that, assuming $H_0$ = 67.8 km s$^{-1}$ Mpc$^{-1}$, $\Omega_M$ = 0.308 and $\Omega_\Lambda$ = 0.692, the redshift provided by NED corrected to the reference frame defined by the 3K cosmic microwave background (CMB; $z$ = 0.003676) results in a luminosity distance of 16.3 Mpc, which is very close to the (redshift-independent) distance assumed in this work. NGC 1448 is part of the sample of the Deep IFS View of Nuclei of Galaxies (DIVING$^{3D}$) survey (Steiner et al., in preparation), which is being conducted by our research group. The goal of this survey is to study, using optical 3D spectroscopy, the central regions of all galaxies in the Southern hemisphere with B < 12.0 and |$b$| < 15\degr. 

There is not much information in the literature about NGC 1448. \citet{con96}, using Very Large Array observations, obtained a flux density of 95.5 mJy at 1.49 GHz. The nuclear emission-line spectrum of this galaxy was originally classified as characteristic of an H II region, based on the H$\alpha$/[N \textsc{ii}] ratio, by \citet{ver86}. However, the detection of the [Ne \textsc{v}]$\lambda14.32, \lambda24.32$ $\mu$m and [O \textsc{iv}]$\lambda$25.89 $\mu$m mid-infrared (MIR) emission lines \citep{gou09,per10} revealed the presence of an AGN, as such high-ionization MIR lines cannot be produced by photoionization by young stars. The mass of the central SMBH in NGC 1448 was estimated using two different techniques. The first involved a relation between the bulge luminosity of the galaxy in the K band and the SMBH mass, which resulted in $log(M_{BH}) = 5.99^{+0.11}_{-0.52}$ M$_{\sun}$ \citep{gou10}. The second technique involved a relation between the pitch angle of the galaxy and the SMBH mass, resulting in $log(M_{BH}) = 7.70 \pm 0.20$ M$_{\sun}$ \citep{dav14}. \citet{tsa15} estimated the star formation rate density in the central kiloparsec of this galaxy, using the luminosity at 8 $\mu$m, and obtained a value of -0.89 M$_{\sun}$ yr$^{-1}$ kpc$^{-2}$.

\begin{figure*}
\begin{center}
   \includegraphics[scale=0.62]{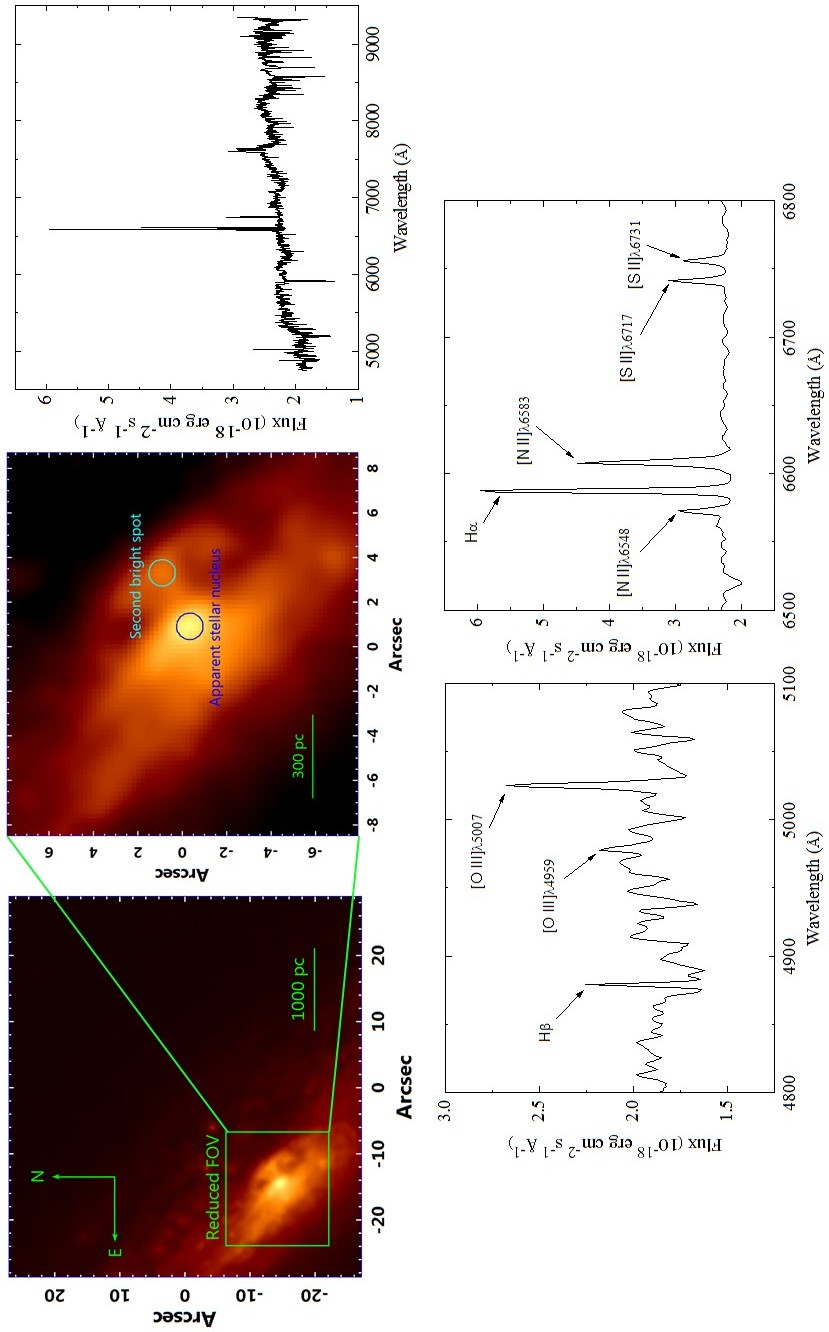} 
  \caption{Top left: image of the data cube of NGC 1448, retrieved from the MUSE data archive, collapsed along its spectral axis. Top middle: the same as on the top left, but with the reduced FOV that is analysed in this work. The blue circle represents the apparent stellar nucleus of the galaxy and the cyan circle corresponds to a second bright spot detected in the data cube with the reduced FOV. Top right: average spectrum of the MUSE data cube with the reduced FOV. Bottom: magnifications of the average spectrum are shown on the top right.}\label{fig01}
\end{center}
\end{figure*}

The only detailed study of the central region of NGC 1448, so far, was performed by \citet{ann17}, using X-ray data obtained with the Nuclear Spectroscopic Array (NuSTAR) and with the Chandra Space Telescope. The authors also analysed optical spectra obtained with the New Technology Telescope (NTT) of the European Southern Observatory (ESO) and high spatial resolution MIR images obtained with the Thermal-Region Camera Spectrograph (T-ReCS) at the Gemini-South telescope. The X-ray spectra modelling revealed that the AGN in this galaxy is actually Compton thick, with a column density along the line of sight of $N_H \geq 2.5 \times 10^{24} cm^{-2}$. The models indicated the presence of a highly inclined torus causing the obscuration. The authors observed that the position of the AGN, determined from the Chandra data, does not coincide with the peak of the stellar emission, which was interpreted as the result of the extinction caused by a dust lane along the line of sight of the AGN (corresponding to the true nucleus), similarly to the cases observed by \citet{pri14}. An emission-line diagnostic diagram analysis \citep{bal81,vei87} from the optical spectroscopic data revealed that the nuclear emission-line spectrum of NGC 1448 is characteristic of a type 2 Seyfert galaxy.

In this work, we present the analysis of an optical data cube of the central region of NGC 1448, obtained with the Multi Unit Spectroscopic Explorer (MUSE; \citealt{bac10}). We focus our analysis on scientific aspects different from the ones addressed by \citet{ann17}. In particular, we analyse the spatial morphology of the line-emitting regions and also the stellar and gas kinematics. The paper is organized as follows. In Section~\ref{sec2}, we describe the data and also the data treatment procedures we applied. In Section~\ref{sec3}, we present the results from the analysis of the line-emitting regions. In Section~\ref{sec4}, we perform stellar archaeology based on the results from a spectral synthesis. In sections~\ref{sec5} and~\ref{sec6}, we analyse the stellar and gas kinematics, respectively. Finally, we discuss our results in Section~\ref{sec7} and present our conclusions in Section~\ref{sec8}. 

\section{Observations and data treatment}\label{sec2}

\begin{figure*}
\begin{center}
   \includegraphics[scale=0.37]{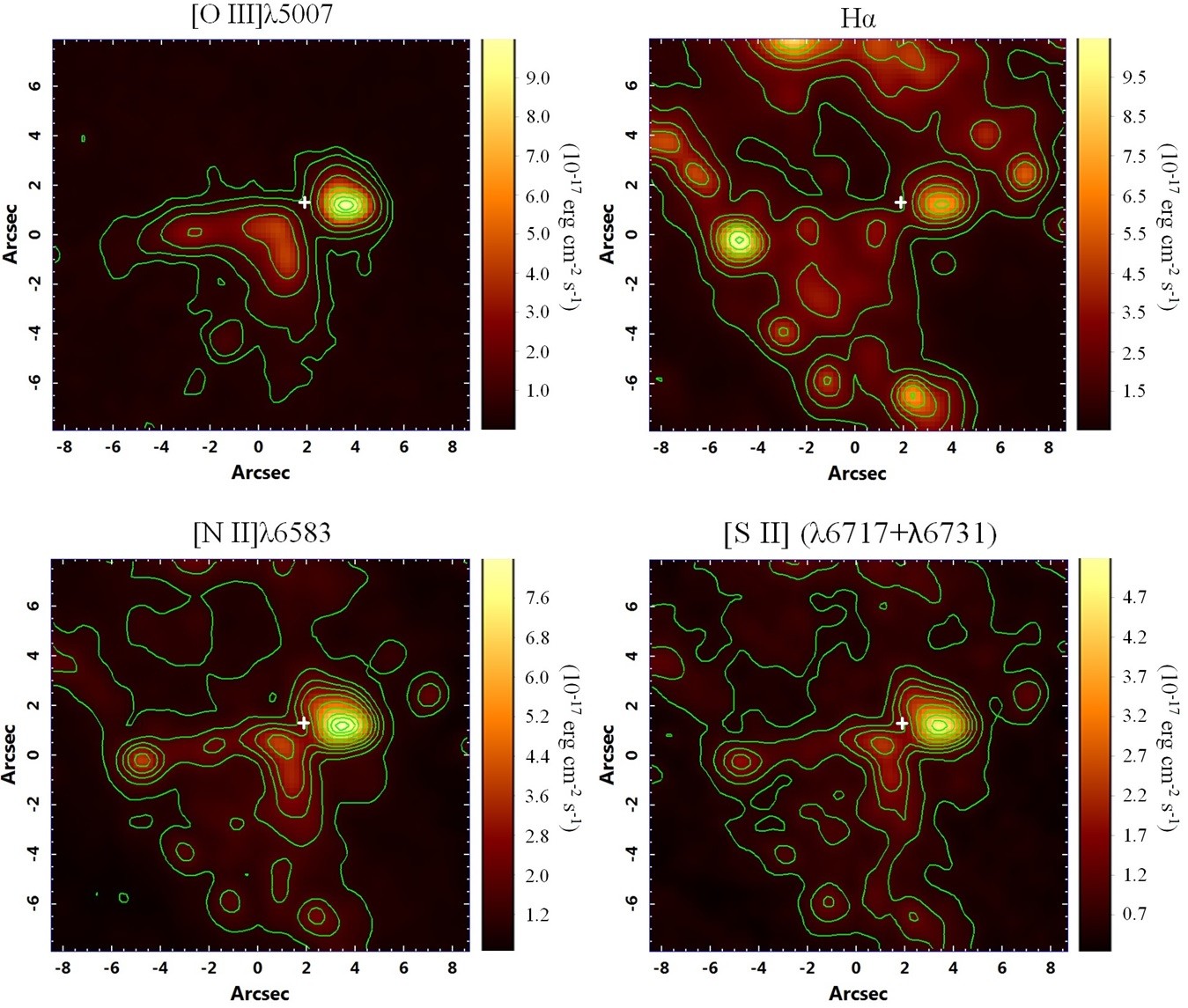} 
  \caption{Integrated flux images of the [O \textsc{iii}]$\lambda$5007, H$\alpha$, [N \textsc{ii}]$\lambda$6583 and [S \textsc{ii}]($\lambda$6717 + $\lambda6731$) emission lines obtained from the gas data cube of NGC 1448. The isocontours are shown in green and the AGN position, provided by the hard X-ray emission (in the range of 2 - 10 keV) detected in the Chandra data of this galaxy, is marked with a white cross in each image.}\label{fig02}
\end{center}
\end{figure*}

An optical data cube of the central region of NGC 1448 was originally obtained, for the DIVING$^{3D}$ survey, with the SOAR Integral Field Spectrograph (SIFS), as part of a science verification program. However, considering that the MUSE data cube of this galaxy has a higher spatial resolution and also spectra with higher signal-to-noise ratio (S/N), we opted to analyse the later in this work. The reduced MUSE data cube of NGC 1448 was retrieved from the ESO data archive. The observations were made on 2017 July 24. The program of the observations is 097.D-0408 (PI: Anderson, Joseph), which is part of the All-weather MUSE Supernova Integral field Nearby Galaxies (AMUSING) Survey III \citep{gal16}. A single 2164 s exposure was taken. The data cube has a spectral coverage of 4750 - 9350 \AA, with a spectral resolution of $R \sim 3027$, and is sampled by spatial pixels (spaxels) of $0.2$ arcsec. The full width at half-maximum (FWHM) effective spatial resolution is $1.18$ arcsec and the field of view (FOV) of the data cube has $1.08$ $\times$ $1.11$ arcmin$^2$.

Since we are using the MUSE data cube of NGC 1448 with the specific purpose of analysing the central region of this galaxy, we resized the data cube. The final FOV, around the observed stellar nucleus of the galaxy, has $17.2$ $\times$ $15.8$ arcsec$^2$ (which is equivalent to $1.37$ $\times$ $1.26$ kpc$^2$, considering a distance of $\sim 16.4$ Mpc). We applied a Butterworth spatial filtering \citep{gon02} to each image of the data cube, in order to remove high spatial frequency noise. This procedure is part of the treatment methodology discussed in \citet{men14,men15,men19}. 

Fig.~\ref{fig01} shows images of the data cube of NGC 1448, collapsed along the spectral axis, with its original FOV and also with the reduced FOV that is analysed in this work. The average spectrum of the data cube with the reduced FOV is also shown. The bright spot corresponding to the apparent stellar nucleus can be seen very close to a darker area with a morphology similar to a ``stripe'' crossing the region. The image also shows, northwest of the apparent stellar nucleus, a second bright spot, in the area corresponding to the ``dark stripe''. This peculiar spatial morphology is probably a consequence of the interstellar extinction caused by a dust lane in this region (as established by \citealt{ann17}). This topic will be discussed in further detail throughout the paper. The average spectrum is relatively noisy, but the H$\beta$, [O \textsc{iii}]$\lambda4959; \lambda$5007, [N \textsc{ii}]$\lambda6548; \lambda6583$, H$\alpha$ and [S \textsc{ii}]$\lambda6717; \lambda6731$ emission lines can be easily detected.

In order to perform a multiwavelength analysis of the central region of NGC 1448, we retrieved X-ray data of this galaxy, taken with the Chandra Space Telescope, from the Chandra public data archive. One 52.6 s exposure was taken with the Advanced CCD Imaging Spectrometer-Spectroscopic (ACIS-S) instrument, on 2014 March 9, as part of the program 15332 (PI: Murray). Based on the event file of the data, we obtained images of NGC 1448 in different wavelengths (energies), in units of counts s$^{-1}$ and with pixels of 0.49 arcsec. These images were combined in a data set with two spatial dimensions and one spectral dimension, analogous to a ``data cube''. Such a data cube was spatially resampled, in order to obtain pixels with a quarter of the original size ($\sim 0.12$ arcsec). We also applied a Butterworth spatial filtering to each image of the resulting data cube, to remove high spatial frequency components, some of which were probably introduced by the spatial resampling procedure. These same Chandra data were previously analysed by \citet{ann17}, who detected a point-like source corresponding to the AGN and also two other compact sources, within a radius of 20 arcsec from the AGN, which are probably X-ray binaries. Our main motivations to reanalyse these data are to determine the correct position of the AGN in the MUSE FOV and also to evaluate the presence of possible extended emission around the source representing the AGN. We spatially resized the Chandra data to match the same FOV of the MUSE data cube (for further detail, see Section~\ref{sec3}).

\section{Line-emitting regions}\label{sec3}

\begin{figure*}
\begin{center}
   \includegraphics[scale=0.30]{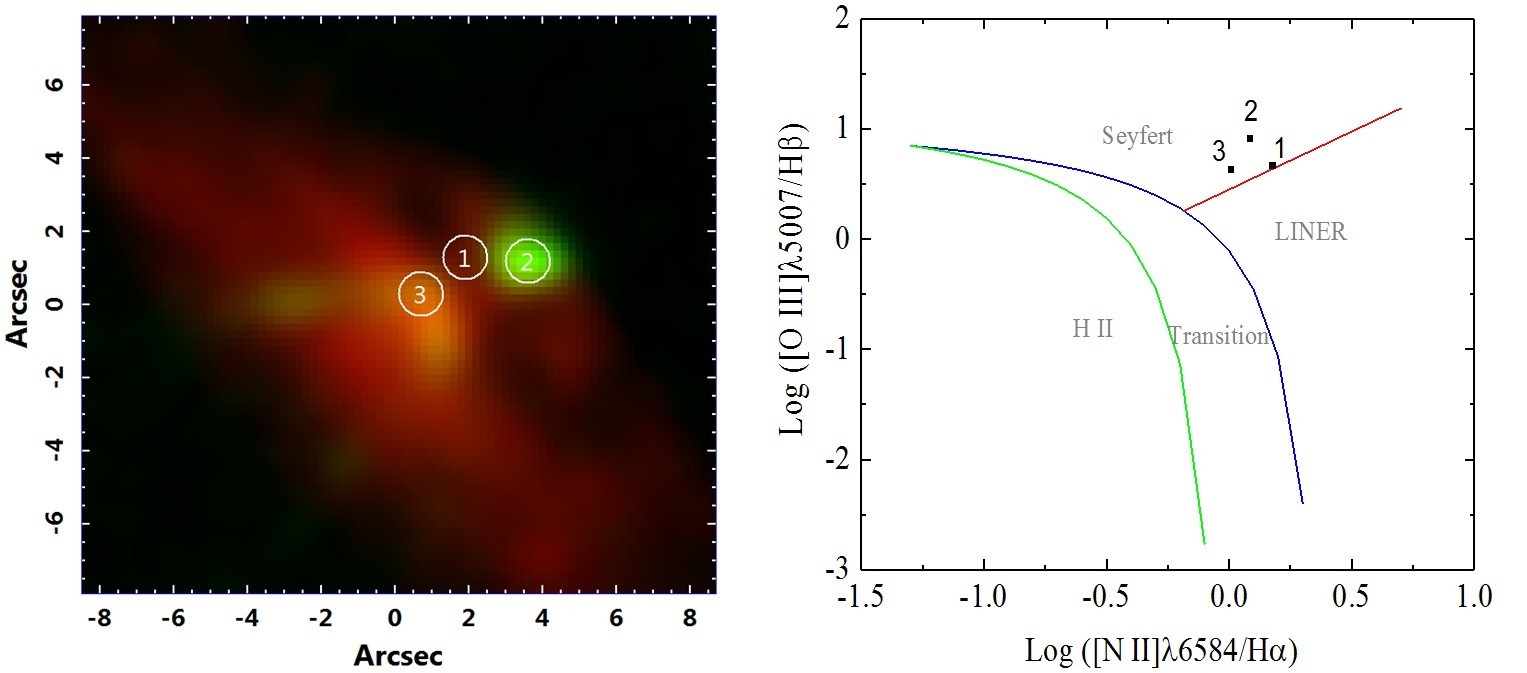} 
  \caption{Left: RG composite image with the red and green colours representing the stellar data cube of NGC 1448, provided by the spectral synthesis, collapsed along its spectral axis and the integrated flux image of the [O \textsc{iii}]$\lambda$5007 emission line, respectively. The circular regions from which the spectra in Fig.~\ref{fig05} were extracted are shown with white circles. Right: diagnostic diagram of [O \textsc{iii}]$\lambda$5007/H$\beta$ $\times$ [N \textsc{ii}]$\lambda$6583/H$\alpha$ with the points representing regions 1, 2 and 3. The blue curve represents the maximum limit for the ionization by a starburst obtained by \citet{kew01}, the green curve corresponds to the empirical division between H II regions and AGNs determined by \citet{kau03} and the red line corresponds to the division between low ionization nuclear emission-line regions (LINERs) and Seyferts determined by \citet{sch07}.}\label{fig03}
\end{center}
\end{figure*}

\begin{figure*}
\begin{center}
   \includegraphics[scale=0.225]{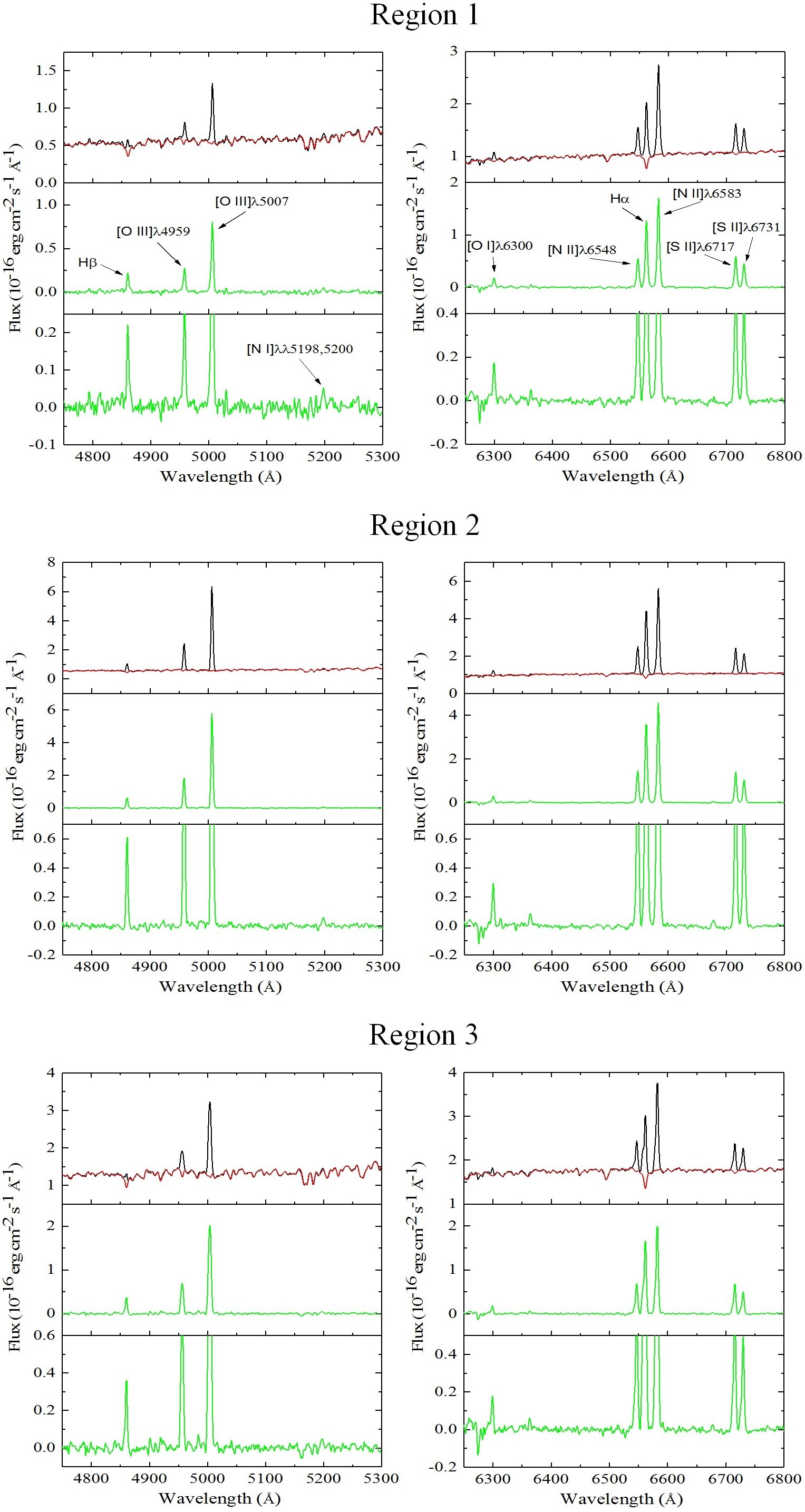} 
  \caption{Spectra extracted from regions 1, 2 and 3 (see Fig.~\ref{fig03}) of the data cube of NGC 1448 before the starlight subtraction. The fits provided by the spectral synthesis are shown in red and the fit residuals are shown in green.}\label{fig04}
\end{center}
\end{figure*}

\begin{table*}
\centering
\caption{Projected distances and position angles of the lines connecting the centres of regions 1, 2 and 3. The projected distances in pc were determined assuming a distance of 16.4 Mpc for NGC 1448.\label{tbl1}}
\begin{tabular}{ccc}
\hline
Regions & Projected distance between the regions & Position angle of the line connecting the regions ($\degr$) \\ 
\hline
1 and 2 & $1.70 \pm 0.22$ arcsec or $135 \pm 17$ pc & $86 \pm 7$ \\
2 and 3 & $2.95 \pm 0.14$ arcsec or $235 \pm 11$ pc & $-61 \pm 5$ \\
1 and 3 & $1.75 \pm 0.22$ arcsec or $139 \pm 17$ pc & $-32 \pm 7$ \\

\hline
\end{tabular}
\end{table*}

\begin{figure*}
\begin{center}
   \includegraphics[scale=0.42]{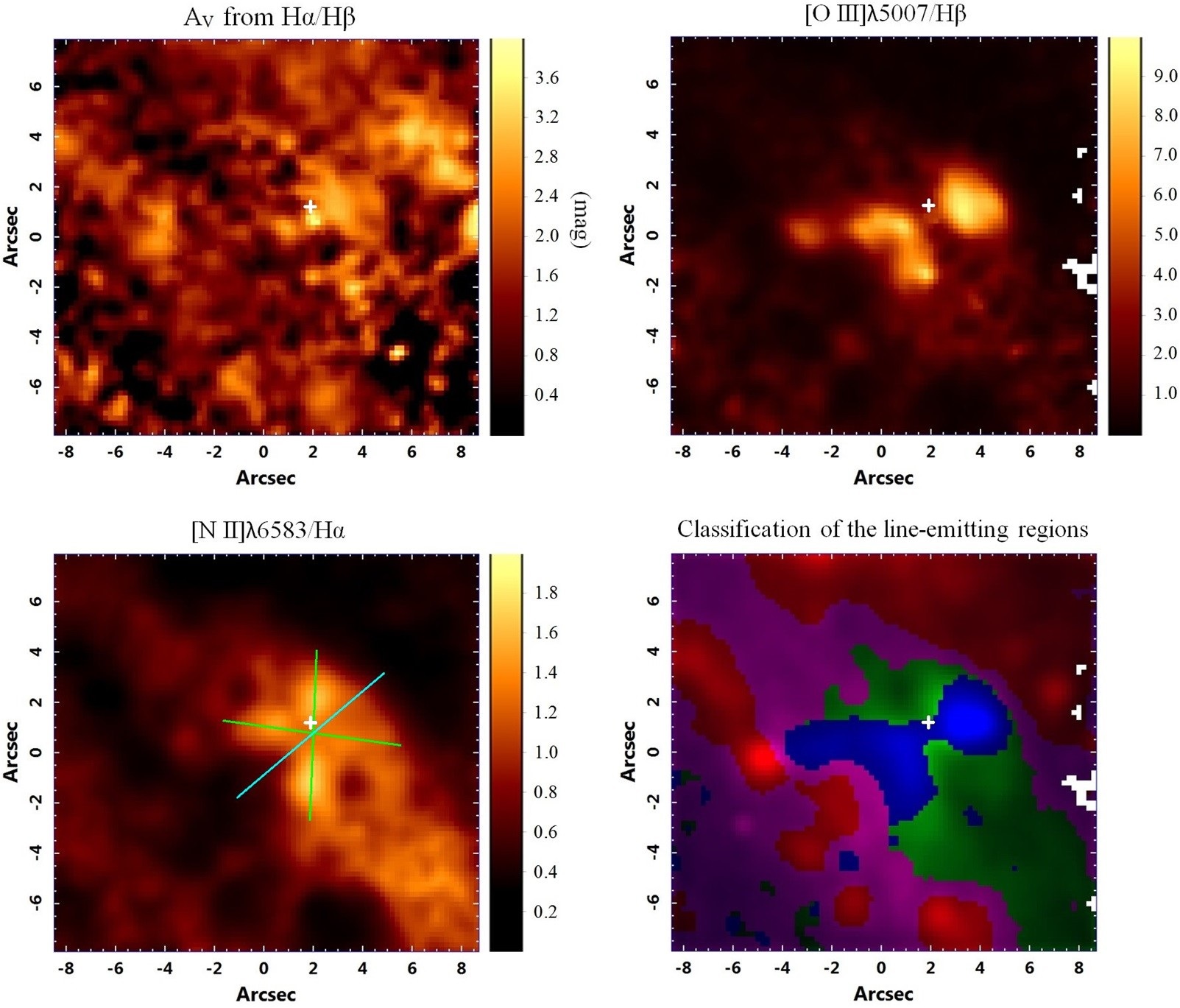} 
  \caption{Top left: $A_V$ map obtained from the values of the Balmer decrement (H$\alpha$/H$\beta$) in the gas data cube of NGC 1448. Top right: map of the [O \textsc{iii}]$\lambda$5007/H$\beta$ ratio obtained from the gas data cube of NGC 1448, after the correction for the interstellar extinction. Bottom left: map of the [N \textsc{ii}]$\lambda$6583/H$\alpha$ ratio obtained from the gas data cube of NGC 1448, after the correction for the interstellar extinction. The green lines represent the walls of the probable ionization cones of the AGN and the cyan line represents the axis of the ionization cones. Bottom right: map showing the classification of the emission-line spectra of the different spatial regions along the FOV of the data cube. The red, magenta, green, and blue colours represent regions with emission-line spectra typical of H II regions, transition objects, LINERs, and Seyfert galaxies, respectively. The white cross in the maps represents the AGN position provided by the hard X-ray emission detected in the Chandra data of this galaxy. The white areas in the maps were masked because the amplitude-to-noise (A/N) ratio of one or more emission lines used for the analysis in such regions was lower than 3.}\label{fig05}
\end{center}
\end{figure*}

\begin{figure*}
\begin{center}
   \includegraphics[scale=0.30]{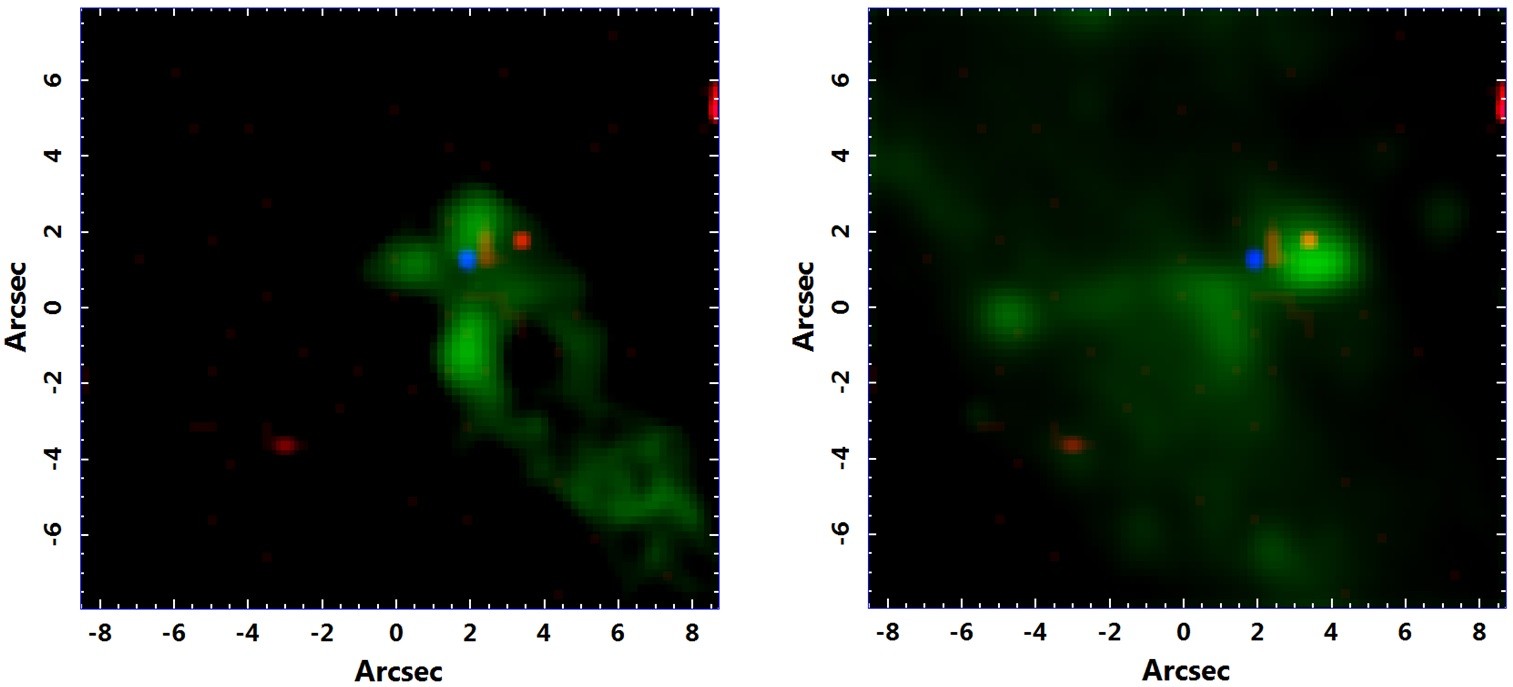} 
  \caption{Left: RGB composite image with the blue and red colours representing the hard and soft (with energies below 2 keV) X-ray emission from the Chandra data, respectively, and the green colour representing  the map of the [N \textsc{ii}]$\lambda$6583/H$\alpha$ ratio from the MUSE gas data cube. Right: RGB composite image with the blue and red colours representing the hard and soft X-ray emission from the Chandra data, respectively, and the green colour representing the integrated flux map of the MUSE gas data cube.}\label{fig06}
\end{center}
\end{figure*}

The stellar continuum subtraction of the spectra in the MUSE data cube of NGC 1448, which is a necessary step for a reliable analysis of the line-emitting regions, was performed with a spectral synthesis. Such a procedure corresponds to a fit of the stellar spectrum of an object with a combination of template spectra from a base. We applied the spectral synthesis to all the spectra of the MUSE data cube of NGC 1448 with the \textsc{starlight} software \citep{cid05}, which, besides combining the template spectra, also convolves them with a Gaussian function, in order to reproduce the widths of the observed absorption lines. Before this procedure, all the spectra of the data cube were corrected for the Galactic extinction, assuming $A_V$ = 0.039 mag (NED) and the extinction law of \citet{car89}, passed to the rest frame, assuming $z$ = 0.00390 (NED), and spectrally resampled with $\Delta\lambda$ = 1\AA. The base of stellar population spectra used for the spectral synthesis was obtained from the Medium-resolution Isaac Newton Telescope Library of Empirical Spectra (MILES; \citealt{san06}), with ages between 1.0 Myr and 13 Gyr and metallicities between $Z = 0.0001$ and $0.05$ ($Z_{\sun} = 0.02$ being the solar metallicity), and is known as CB2007. The spectral synthesis provides, among other things, a synthetic stellar spectrum corresponding to the fit obtained by the procedure. Such synthetic spectra were subtracted from the observed ones, resulting in a data cube (from now on, the ``gas data cube'') containing, essentially, emission lines.

\begin{table*}
\centering
\caption{Emission-line ratios, extinction values ($A_V$), and luminosities of the H$\alpha$ emission line ($L_{H\alpha}$) obtained for the spectra of the three line-emitting regions in the data cube of NGC 1448. The values of $L_{H\alpha}$ were calculated, after the correction for the interstellar extinction, assuming a distance of 16.4 Mpc for NGC 1448 and its uncertainties were estimated taking into account only the uncertainties of the integrated fluxes of the H$\alpha$ emission lines.\label{tbl2}}
\begin{tabular}{cccccccc}
\hline
Region & $\frac{[O\textsc{iii}]\lambda5007}{H\beta}$ & $\frac{[N\textsc{ii}]\lambda6583}{H\alpha}$ & $\frac{[S\textsc{ii}](\lambda6716 + \lambda6731)}{H\alpha}$ & $\frac{[O \textsc{i}]\lambda6300}{H\alpha}$ & $\frac{[S \textsc{ii}\lambda6717}{[S \textsc{ii}]\lambda6731}$ & $A_V$ (mag) & $L_{H\alpha}$ ($10^{38}$ erg s$^{-1}$) \\ 
\hline
Region 1 (AGN) & $4.63 \pm 0.24$ & $1.50 \pm 0.04$ & $0.84 \pm 0.04$ & $0.165 \pm 0.025$ & $1.23 \pm 0.04$ & $2.68 \pm 0.18$ & $1.64 \pm 0.03$ \\
Region 2       & $8.17 \pm 0.09$ & $1.21 \pm 0.03$ & $0.62 \pm 0.03$ & $0.084 \pm 0.020$ & $1.212 \pm 0.019$ & $2.40 \pm 0.06$ & $3.48 \pm 0.05$ \\
Region 3       & $4.28 \pm 0.12$ & $1.02 \pm 0.03$ & $0.52 \pm 0.04$ & $0.081 \pm 0.021$ & $1.35 \pm 0.08$ & $0.70 \pm 0.10$ & $0.572 \pm 0.010$  \\

\hline
\end{tabular}
\end{table*}

Fig.~\ref{fig02} shows integrated flux images of the [O \textsc{iii}]$\lambda$5007, H$\alpha$, [N \textsc{ii}]$\lambda$6583 and [S \textsc{ii}]($\lambda$6717 + $\lambda6731$) emission lines of the gas data cube. The AGN position, as revealed by the compact hard X-ray (in the range of 2 - 10 keV) source detected in the Chandra data, was marked with a cross in each image in Fig.~\ref{fig02}. The [O \textsc{iii}]$\lambda$5007 image shows a triangular-shaped structure south-east of the AGN. This feature is also seen in the [N \textsc{ii}]$\lambda$6583, [S \textsc{ii}]($\lambda$6717 + $\lambda6731$), and H$\alpha$ images, although it is not so well defined in the last one. The H$\alpha$ image shows a number of H II regions around the nucleus. The brightest line-emitting region, west of the AGN, without taking into account the H II regions, is compact and not coincident with either the AGN or the apparent stellar nucleus. From now on, we will refer to such region simply as the brightest line-emitting region. The separation between the apparent stellar and the brightest line-emitting region is consistent with the findings of \citet{ann17}. However, the fact that the AGN position does not coincide with the brightest line-emitting region will be discussed in further detail in Section~\ref{sec7}. Based on the spatial morphology of the triangular-shaped feature in the images in Fig.~\ref{fig02}, we propose that it is part of the ionization cone of the AGN in NGC 1448. The brightest line-emitting region, west of the AGN, on the opposite side is probably part of a second ionization cone, as discussed below. 

In order to obtain representative emission-line spectra of the AGN and of circumnuclear regions, and also to show the quality of the stellar continuum subtraction, we extracted the spectra from three circular regions, centred at the AGN (Region 1) and at two positions in the probable ionization cones on both sides of the AGN (regions 2 and 3). The radius of each extraction region was taken as half of the FWHM of the point spread function (PSF). Fig.~\ref{fig03} shows the positions of the three extraction regions. The centre of Region 2 coincides with the brightest line-emitting region (defined before), west of the AGN, while the centre of Region 3 coincides with the apparent stellar nucleus. The spectra extracted from regions 1, 2 and 3 are shown in Fig.~\ref{fig04}, together with the fits obtained with the spectral synthesis and also the fit residuals, which correspond to the emission-line spectra of these three areas. Table~\ref{tbl1} shows the projected distances and also the position angles of the lines connecting the centres of these regions.

The [N \textsc{ii}] and H$\alpha$ emission lines in the gas data cube of NGC 1448 are not significantly blended (as can be seen in the extracted spectra shown in Fig.~\ref{fig04}). Therefore, in order to perform a diagnostic diagram analysis of the spectra of regions 1, 2, and 3, the integrated fluxes of the [N \textsc{ii}]$\lambda$6583 and H$\alpha$ emission lines, and also of the H$\beta$ and [O \textsc{iii}]$\lambda$5007 emission lines, were determined via direct integration. First, the emission-line spectra of regions 1, 2 and 3 were corrected for the interstellar extinction, using the obtained values of the Balmer decrement (H$\alpha$/H$\beta$) and assuming an intrinsic Balmer decrement of 3.10 (appropriate for AGNs; \citealt{ost06}) and an extinction law of \citet{car89}. Then we calculated the integrated fluxes of the H$\beta$, [O \textsc{iii}]$\lambda$5007, H$\alpha$, and [N \textsc{ii}]$\lambda$6583 emission lines, and the [O \textsc{iii}]$\lambda$5007/H$\beta$ and [N \textsc{ii}]$\lambda$6583/H$\alpha$ emission line ratios. The uncertainties of the integrated fluxes were calculated taking into account the spectral noise and also the fit residuals of the spectral synthesis. The uncertainties of the emission-line ratios were calculated by propagation of the uncertainties of the integrated fluxes. Fig.~\ref{fig03} shows the diagnostic diagram of [O \textsc{iii}]$\lambda$5007/H$\beta$ $\times$ [N \textsc{ii}]$\lambda$6583/H$\alpha$ with the points corresponding to regions 1, 2, and 3. The three points fall in the Seyfert galaxies branch, which is consistent with the previous results obtained by \citet{ann17}, using optical long-slit spectra. Since no broad component was detected in the observed emission lines, the most appropriate classification for the AGN in NGC 1448 is Seyfert 2. The results also show that Region 2 has a relatively higher ionization degree than the other two and the point corresponding to Region 1 is actually very close to the dividing line between the Seyferts and low ionisation nuclear emission-line regions (LINERs) in the diagnostic diagram. Table~\ref{tbl2} shows the emission-line ratios obtained for the spectra of regions 1, 2, and 3 (the values of the [S\textsc{ii}]($\lambda$6716 + $\lambda$6731)/H$\alpha$, [O \textsc{i}]$\lambda$6300/H$\alpha$, and [S \textsc{ii}]$\lambda$6717/[S \textsc{ii}]$\lambda$6731 emission-line ratios were also included), together with the $A_V$ values determined from the Balmer decrements and the luminosities of the H$\alpha$ emission line ($L_{H\alpha}$). The $A_V$ values determined for regions 1 and 2 are compatible, at the 2$\sigma$ level, and are significantly higher and not compatible, even at the 3$\sigma$ level, with the $A_V$ value obtained for Region 3, indicating the presence of a higher interstellar extinction, probably due to a dust lane, along the lines of sight of regions 1 and 2.

One of the main advantages of good-quality 3D spectroscopy, in comparison with long-slit spectroscopy, is the possibility to analyse the emission-line spectra of different spatial regions, with a high spatial resolution. Therefore, we repeated the diagnostic diagram analysis described before for all the emission-line spectra in the gas data cube of NGC 1448. First, we constructed an $A_V$ map, based on the values of the Balmer decrement of each spectrum of the data cube and assuming an extinction law of \citet{car89}. We also assumed an intrinsic Balmer decrement of 2.86 for the spectra characteristic of H II regions and of 3.10 for the other spectra \citep{ost06}. Using the obtained $A_V$ values, we corrected all the spectra of the gas data cube for the interstellar extinction. Then, we obtained, via direct integration, the integrated fluxes of the H$\beta$, [O \textsc{iii}]$\lambda$5007, H$\alpha$, and [N \textsc{ii}]$\lambda$6583 emission lines for each spectrum and constructed the maps of the [O \textsc{iii}]$\lambda$5007/H$\beta$ and [N \textsc{ii}]$\lambda$6583/H$\alpha$ emission line ratios. Using such maps, based on the [O \textsc{iii}]$\lambda$5007/H$\beta$ $\times$ [N \textsc{ii}]$\lambda$6583/H$\alpha$ diagnostic diagram, we could classify the spectrum of each spaxel of the gas data cube as being characteristic of an H II region, a transition object, a Seyfert galaxy, or a LINER. Fig.~\ref{fig05} shows the maps of $A_V$, [O \textsc{iii}]$\lambda$5007/H$\beta$, [N \textsc{ii}]$\lambda$6583/H$\alpha$, and the classification of the line-emitting regions. 

The $A_V$ map in Fig.~\ref{fig05} is considerably noisy but reveals areas with higher interstellar extinction northwest of the apparent stellar nucleus, which is consistent with the hypothesis of a dust lane in such region. The spatial morphologies of the [O \textsc{iii}]$\lambda$5007/H$\beta$ and [N \textsc{ii}]$\lambda$6583/H$\alpha$ maps are compatible with the presence of two ionization cones around the central AGN (one towards south-east and the other towards northwest of the AGN). Actually, the [N \textsc{ii}]$\lambda$6583/H$\alpha$ map shows in more detail the walls of the two ionization cones. This is expected, as the values of the [N \textsc{ii}]$\lambda$6583/H$\alpha$ ratio should be, in principle, higher in the partially ionized region, which is more significant at the borders of an ionization cone. Based on the [N \textsc{ii}]$\lambda$6583/H$\alpha$ map, we estimated a position angle of $PA_{cones} = -50\degr \pm 7\degr$ for the axis of the ionization cones of the AGN in NGC 1448.

The map showing the classification of the line-emitting regions in Fig.~\ref{fig05} reveals that the areas along the probable ionization cones have emission-line spectra characteristic of Seyfert galaxies, which is consistent with the results provided by the diagnostic diagram in Fig.~\ref{fig03}. Regions located perpendicularly to the ionization cones show emission-line spectra typical of LINERs. Farther from the AGN we can see a number of H II regions, as already inferred from the integrated flux map of H$\alpha$ in Fig.~\ref{fig02}, and also areas with spectra characteristic of transition objects.

The RGB composite images in Fig.~\ref{fig06} show comparisons between the hard X-ray emission, the soft X-ray emission (with energies below 2 keV), and the maps of the [N \textsc{ii}]$\lambda$6583/H$\alpha$ ratio and of the integrated flux of the entire emission-line spectra. Due to the low S/N ratio of the Chandra data, one must be cautious about these results. Nevertheless, while the hard X-ray emission is apparently point-like, the soft X-ray emission is not. The soft X-ray emission is also not centred at the position of the AGN and may even be extended, towards northwest, coinciding with the region corresponding to one of the probable ionization cones of the AGN. The area with soft X-ray emission is nearly coincident with the brightest line-emitting region (Region 2). Considering that, we believe that such a soft X-ray emission may be associated with one of the ionization cones of the AGN in NGC 1448.

\section{Stellar archaeology}\label{sec4}

\begin{figure*}
\begin{center}
   \includegraphics[scale=0.38]{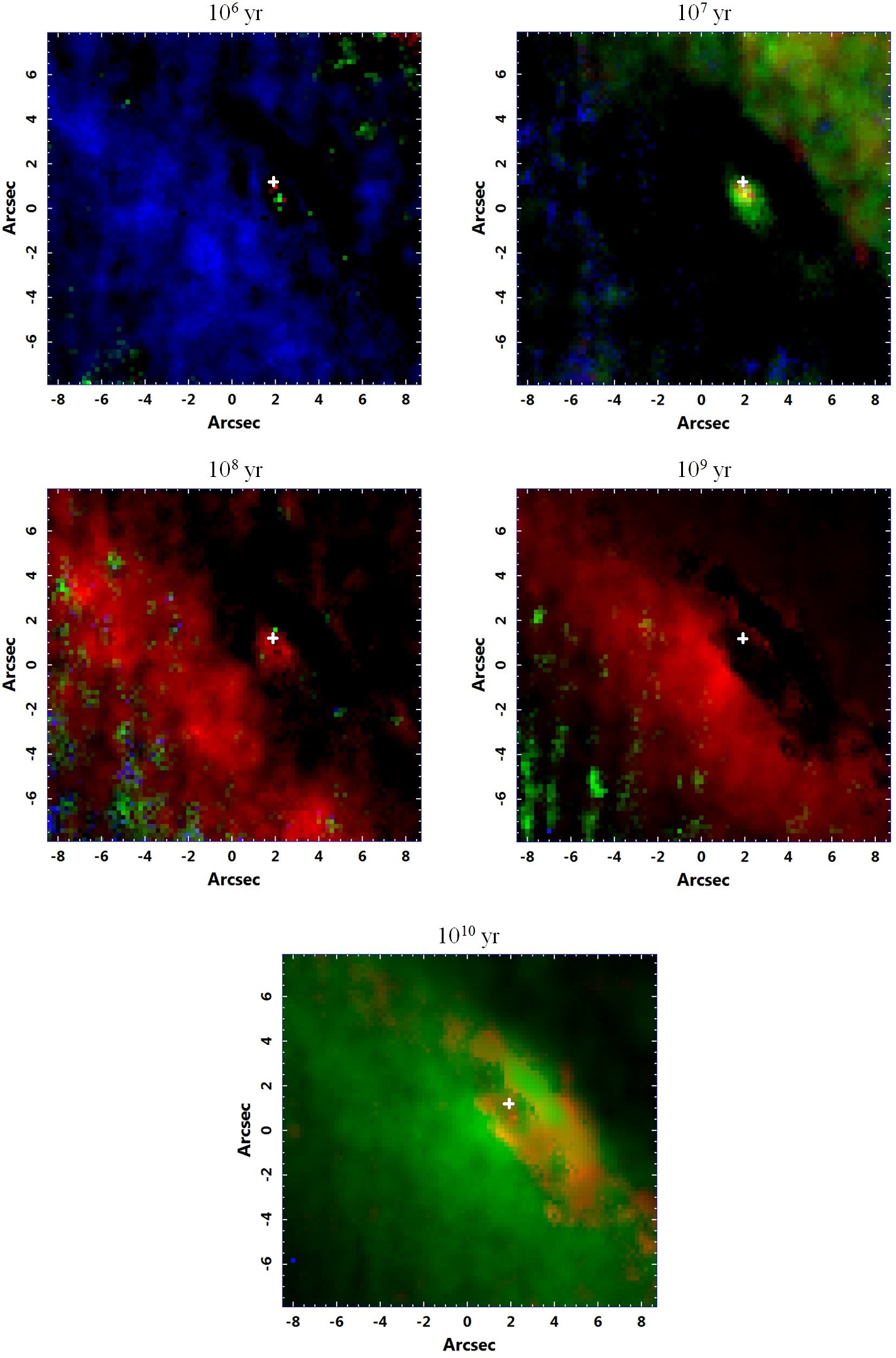} 
  \caption{Integrated flux maps of the stellar populations detected by the spectral synthesis applied to the MUSE data cube of NGC 1448, with ages of $10^6$, $10^7$, $10^8$, $10^9$, and $10^{10}$ yr. The blue, green, and red colours represent low ($Z = 0.0001$ and $0.0004$), intermediate ($Z = 0.004$ and $0.008$) and high ($Z = 0.02$ and $0.05$) metallicities ($Z_{\sun} = 0.02$ being the solar metallicity), respectively. The white cross in the maps represents the AGN position provided by the hard X-ray emission detected in the Chandra data of this galaxy.}\label{fig07}
\end{center}
\end{figure*}

\begin{figure}
\begin{center}
   \includegraphics[scale=0.25]{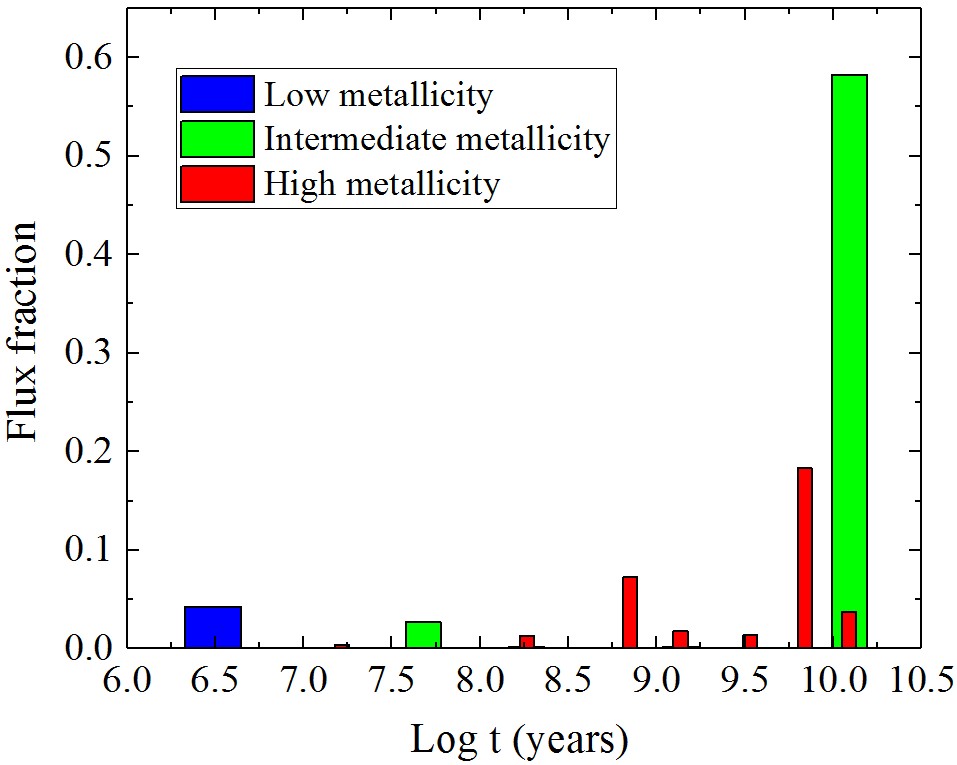} 
  \caption{Histogram of the flux fractions associated with all stellar populations taken into account in the spectral synthesis applied to the MUSE data cube of NGC 1448, with reduced FOV.}\label{fig08}
\end{center}
\end{figure}

The spectral synthesis performed with the \textsc{starlight} software provides, in addition to a synthetic stellar spectrum, different parameters, such as the flux fraction associated with each stellar population taken into account in the fit and the interstellar extinction in the observed galaxy. Since this procedure was applied to the spectrum corresponding to each spaxel of the MUSE data cube of NGC 1448 (with reduced FOV), we could construct maps of these parameters. Fig.~\ref{fig07} shows flux maps of stellar populations with ages of $10^6$, $10^7$, $10^8$, $10^9$, and $10^{10}$ yr. We evaluated that a minimum S/N ratio of 10 for the stellar continuum is required, in order to obtain reliable results from the spectral synthesis. Since the S/N ratio values in the wavelength range of $6600 - 6700$\AA~were higher than 10 along the reduced FOV of the MUSE data cube, all the results provided by the spectral synthesis were sufficiently reliable to be included in this work. We also constructed a histogram (Fig.~\ref{fig08}) showing the flux fractions associated with all the stellar populations taken into account in the spectral synthesis. 

The maps in Fig.~\ref{fig07} and the histogram in Fig.~\ref{fig08} reveal that the stellar populations scattered across the reduced MUSE FOV have mainly intermediate ($Z = 0.004$ and $0.008$) and high ($Z = 0.02$ and $0.05$) metallicities. Most of the stellar flux ($\sim 58.2\%$) is due to old stellar populations, with $1.2 \times 10^{10}$ yr and intermediate metallicity. Another significant fraction of the stellar flux ($\sim 18.3\%$) comes from an also old stellar population, with $6.9 \times 10^9$ yr and high metallicity. Besides that, an intermediate age ($7.1 \times 10^8$ yr) stellar population with high metallicity and a young stellar population ($3.1 \times 10^6$ yr) with low metallicity ($Z = 0.0001$ and $0.0004$) contribute with lower, but still significant, flux fractions ($\sim 7.3\%$ and $\sim 4.2\%$, respectively). One interesting point to be mentioned is that the integrated flux image of stellar populations with $10^7$ yr shows an apparent compact cluster around the AGN position (provided by the hard X-ray image from the Chandra data). Although the structure also shows relatively older stars (the cluster is visible in the integrated flux image of the stellar populations with $10^8$ yr), it may be interpreted as a young nuclear stellar cluster in this galaxy. 

\begin{figure}
\begin{center}
   \includegraphics[scale=0.38]{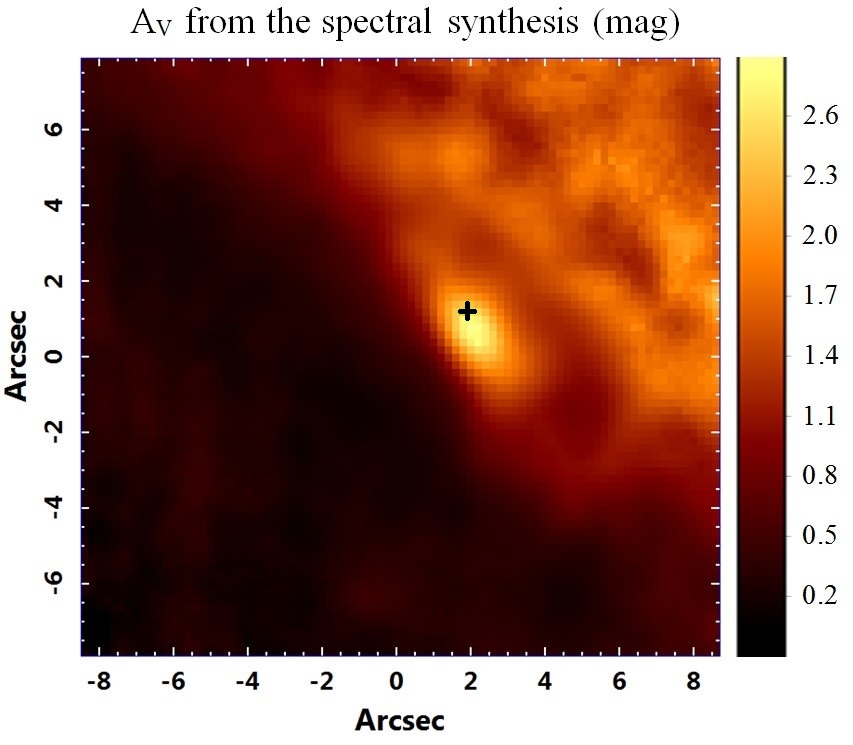} 
  \caption{Map of the values of $A_V$ provided by the spectral synthesis of the data cube of NGC 1448. The black cross represents the AGN position provided by the hard X-ray emission detected in the Chandra data of this galaxy.}\label{fig09}
\end{center}
\end{figure}

Fig.~\ref{fig09} shows the map of the values of $A_V$ provided by the spectral synthesis. It is easy to see that the interstellar extinction is higher along a region located northwest of the apparent stellar nucleus, which suggests the presence of a dust lane. Although the $A_V$ map in Fig.~\ref{fig05} (obtained from the values of the Balmer decrement) is noisier than the one obtained from the spectral synthesis, both show higher values northwest of the AGN. Therefore, the obscuration caused by the dust lane in this region is affecting the stellar and gas emission. 

\begin{figure}
\begin{center}
   \includegraphics[scale=0.38]{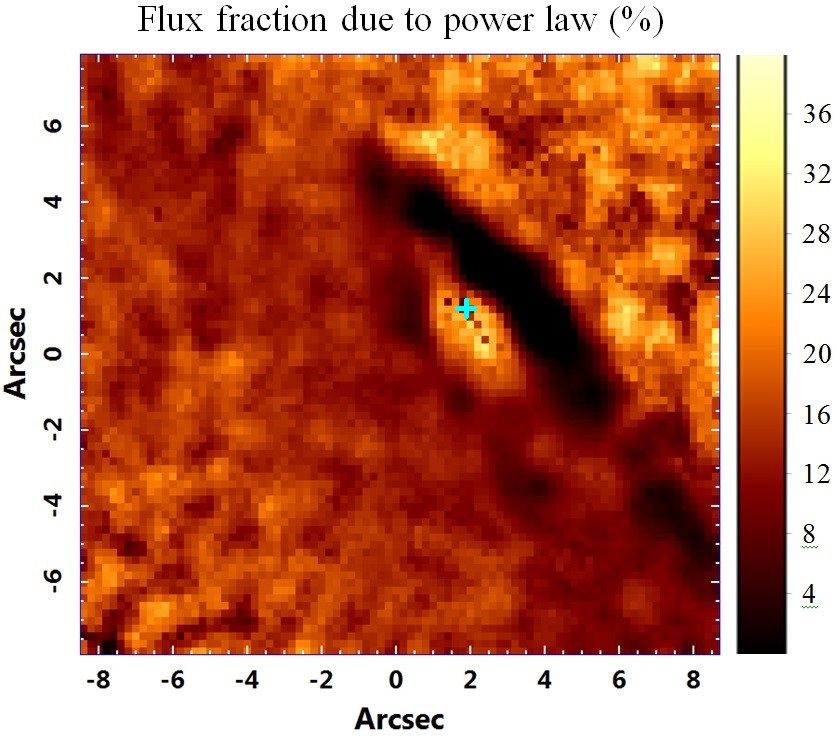} 
  \caption{Map of the flux fractions (\%) associated with the power law, representing the AGN featureless continuum with a spectral index of 1.5, taken into account in the spectral synthesis of the data cube of NGC 1448. The cyan cross represents the AGN position provided by the hard X-ray emission detected in the Chandra data of this galaxy.}\label{fig10}
\end{center}
\end{figure}

It is well known that the spectral synthesis has certain problems that must be taken into account, such as different sources of uncertainties and also degeneracies in the results. Although it is very difficult to determine the actual uncertainties of the results, we applied the procedure described below to the spectra extracted from regions 1, 2, and 3 to, at least, obtain an estimate of these uncertainties. First, for each of these three spectra, we constructed a graph of the cumulative flux fraction as a function of the ages of the stellar populations taken into account for the spectral synthesis and also a graph of the cumulative flux fraction as a function of the metallicities of the stellar populations. Then we determined the age and the metallicity of the stellar populations at which the cumulative flux fractions were equal to $50\%$. We named these parameters as the median age ($T_{med}$) and the median metallicity ($Z_{med}$). We used these same parameters in the analysis of the spectral synthesis in \citet{men18}. The values obtained for regions 1, 2, and 3 were $T_{med}(1) = 1.27 \times 10^{10}$ yr, $Z_{med}(1) = 0.0066$, $T_{med}(2) = 1.3 \times 10^{10}$ yr, $Z_{med}(2) = 0.0061$, and $T_{med}(3) = 1.21 \times 10^{10}$ yr, $Z_{med}(3) = 0.0070$. After that, we constructed histograms of the spectral noise in the residual spectra of regions 1, 2, and 3 (obtained after the subtraction of the synthetic stellar spectra), taking into account only spectral regions without emission lines. We fitted Gaussian functions to the obtained histograms and then generated Gaussian distributions of random noise, with the same widths of the Gaussians fitted to the histograms. Finally, these distributions of random noise were added to the synthetic stellar spectra of regions 1, 2, and 3, the spectral synthesis was applied to the resulting ``noisy'' spectra and the corresponding values of $T_{med}$ and $Z_{med}$ were determined. The uncertainties of $T_{med}$ and $Z_{med}$ obtained for the original spectra of regions 1, 2, and 3 were taken as the standard deviations of all the values calculated for the ``noisy'' spectra. The results were 0.05 and 0.14 dex for $T_{med}$ and $Z_{med}$, respectively, in Region 1, 0.08 and 0.04 dex for $T_{med}$ and $Z_{med}$ in Region 2 and 0.020 and 0.021 dex for $T_{med}$ and $Z_{med}$ in Region 3. The same approach was used to estimate the uncertainties of the $A_V$ values obtained from the spectral synthesis. The results were $A_V(1) = 2.31$ mag, $A_V(2) = 1.41$ mag, and $A_V(3) = 0.821$ mag for the spectra of regions 1, 2, and 3 and the corresponding uncertainties were 0.22, 0.29, and 0.014 mag, respectively. Taking all that into account, we can say that the uncertainties of $T_{med}$ and $Z_{med}$ in the data cube of NGC 1448 are in the ranges of 0.02 - 0.10 dex and 0.02 - 0.15 dex, respectively, and the uncertainties of $A_V$ are in the range of 0.01 - 0.3 mag.

The uncertainties estimated in the previous analysis, however, do not take into account the degeneracies in the results provided by the spectral synthesis. Therefore, one must be cautious with all the ages and metallicities obtained with a spectral synthesis. A possible effect of the degeneracies in this methodology is related to the featureless continuum emitted by AGNs. In certain circumstances (and specific interstellar extinction values), the spectral synthesis may interpret the featureless continuum of an AGN as a stellar emission from young stars and vice versa. In order to evaluate the significance of this effect in the analysis presented in this work, we reapplied the spectral synthesis to the MUSE data cube of NGC 1448, including a power law (representing the AGN featureless continuum), with a spectral index of 1.5, in the base of stellar population spectra. Fig.~\ref{fig10} shows the map of the flux fractions associated with the AGN featureless continuum detected by the spectral synthesis. We can see that a significant contribution from the featureless continuum to the observed spectra was detected in the surroundings of the AGN, where the previous analysis revealed the presence of $10^7$ yr stellar populations with intermediate metallicity (Fig.~\ref{fig07}). However, the featureless continuum was also detected at larger distances from the AGN (where significant contributions from young stellar populations were detected in the previous analysis). Although a collimation and scattering of an AGN featureless continuum may result in the detection of this emission farther from the AGN (e.g. \citealt{men13}), we believe that it is very unlikely that all the featureless continuum identified in Fig.~\ref{fig10} is real, part of it probably corresponding to the emission from young stars. The result in Fig.~\ref{fig10} shows why the parameters provided by the spectral synthesis should be analysed with caution. We opted to take into account in the discussion of this work (see Section~\ref{sec7}) the ages and metallicities provided by spectral synthesis without considering the AGN featureless continuum. Nevertheless, the possible presence of this spectral feature is relevant for the scenario presented here. We also argue that, although the exact values of ages and metallicities shown in Figs.~\ref{fig07} and ~\ref{fig08} may not be precise, since the same procedure (subject to the same degeneracies) was applied to all the spectra (without considering an AGN featureless continuum), differences between the stellar populations detected by the spectral synthesis in different spatial regions of the data cube are probably correct. In addition, it is very unlikely that the procedure cannot differentiate between very young and very old stellar populations (as the corresponding spectral properties are considerably different). In summary, we believe that the spectral synthesis we applied in this work can detect differences between the stellar populations (especially between very young and very old populations) in different spatial regions of the data cube, even if the exact values of the ages and metallicities provided by the procedure are not precise.

\section{Stellar kinematics}\label{sec5}

\begin{figure*}
\begin{center}
   \includegraphics[scale=0.35]{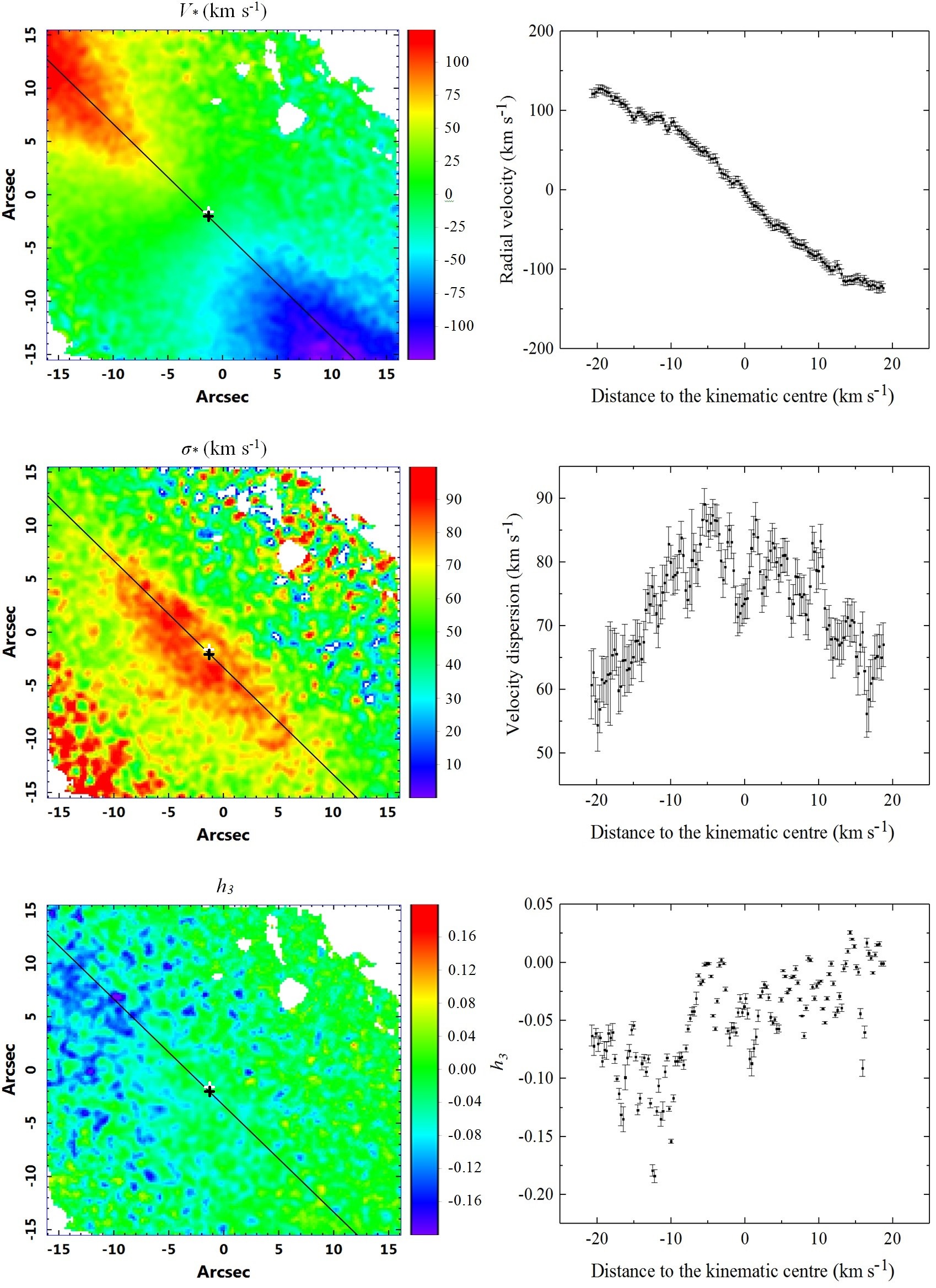} 
  \caption{Maps of the values of (top) the stellar radial velocity ($V_*$), (middle) the stellar velocity dispersion ($\sigma_*$) and (bottom) the Gauss-hermite coefficient $h_3$, obtained with the pPXF technique applied to the MUSE data cube of NGC 1448. The curves on the right were extracted from the line of nodes of the $V_*$ map, which is represented by the black line in the maps. The black cross represents the position of the kinematic centre and the white cross corresponds to the AGN position provided by the hard X-ray emission detected in the Chandra data of this galaxy.}\label{fig11}
\end{center}
\end{figure*}

One of the advantages of using 3D spectroscopy is the possibility of obtaining information about the spatial morphology of the stellar or gas kinematics along a given FOV. In order to obtain the stellar kinematic parameters of the central region of NGC 1448, we applied the Penalized Pixel Fitting (pPXF) procedure \citep{cap04} to the spectrum corresponding to each spaxel of the MUSE data cube. Similarly to the spectral synthesis performed by the \textsc{starlight} software, the pPXF method fits the stellar spectrum of an object with a combination of stellar population spectra from a given base; however, in this case, such template spectra are convolved with a Gauss-Hermite expansion. This technique provides the values of: the stellar radial velocity ($V_*$), the stellar velocity dispersion ($\sigma_*$), and the Gauss-Hermite coefficient $h_3$, which reveals asymmetries in the absorption lines in relation to Gaussian profiles. The pPXF method was applied to the MUSE data cube with the same base of stellar population spectra used to perform the spectral synthesis with the \textsc{starlight} software. Although this work is focused on the study of the central region of NGC 1448, we noticed that more accurate results of the analysis of the stellar kinematics (such as the position angle of the kinematic axis and the coordinates of the kinematic centre) are obtained by applying the pPXF method to a data cube with a somewhat larger FOV than that of the data cube used for the analysis of the line-emitting regions (Section~\ref{sec3}). Therefore, the results discussed in this section were obtained with a data cube with such a larger FOV ($32.2$ $\times$ $31.0$ arcsec$^2$). Fig.~\ref{fig11} shows the maps of $V_*$, $\sigma_*$, and $h_3$ provided by the pPXF technique applied to the MUSE data cube of NGC 1448 (with a larger FOV), together with the curves extracted along the axis corresponding to the line of nodes of the $V_*$ map. It is worth mentioning that the line of nodes represents the intersection between the plane of the observed stellar rotating disc (around the nucleus of NGC 1448) and the plane of the sky. In the $V_*$ map, the line of nodes connects the points with the maximum and minimum observed radial velocity values. We evaluated that, similarly to the case of the spectral synthesis performed with the \textsc{starlight} software, a minimum S/N ratio of 10 is necessary in order to obtain reliable kinematic parameters from the pPXF technique. We calculated the S/N ratio in the wavelength range of $6600 - 6700$\AA~and all the areas with S/N ratio in this spectral region lower than 10 were masked in the maps in Fig.~\ref{fig11}. 

We can see that the $V_*$ map and the $V_*$ curve, extracted along the line of nodes, in Fig.~\ref{fig11} show a pattern consistent with a stellar rotating disc around the nucleus. The kinematic centre in the $V_*$ map was taken as the point, along the line of nodes, where the $V_*$ value was equal to the average between the maximum and minimum values. Such average $V_*$ was actually subtracted from the corresponding map. Therefore, the $V_*$ map in Fig.~\ref{fig11} shows the $V_*$ values relative to the kinematic centre, whose positional uncertainty (1$\sigma$) was estimated as $\sim 0.6$ arcsec. The kinematic centre is not exactly coincident with the AGN position, provided by the hard X-ray emission detected in the Chandra data, but these two positions are compatible, at the 1$\sigma$ level. The amplitude of the $V_*$ curve is $\sim 125$ km s$^{-1}$ and the position angle of the axis corresponding to the line of nodes is $PA_{V_*} = 45\degr \pm 7\degr$. The $\sigma_*$ map is noisier than the $V_*$ map, but it shows that the $\sigma_*$ values in the inner $10$ arcsec of the rotating stellar disc are in the range of $80 - 90$ km s$^{-1}$. In the surroundings of the kinematic centre and of the AGN position, one can see an apparent reduction of the $\sigma_*$ values, which could be characterized as a possible $\sigma$-drop in this galaxy. The $\sigma_*$ curve reveals that the values in the $\sigma$-drop region are actually compatible with the values outside this region, at the $2\sigma$ or $3\sigma$ level. As a consequence, we cannot confirm the presence of this $\sigma$-drop, but the fact that it is located in the same region of the AGN and also in the same region where a possible AGN featureless continuum was detected by the spectral synthesis (see Fig.~\ref{fig10}) suggests that this $\sigma$-drop may be real and associated with the AGN (this topic will be discussed in further detail in Section~\ref{sec7}). The $h_3$ map is certainly the noisiest in Fig.~\ref{fig11}, with certain regions showing no recognizable pattern; however, we can see an apparent anticorrelation with the $V_*$ map, although the values in the $h_3$ curve are not symmetric relative to the kinematic centre. 

The uncertainties of the stellar kinematic parameters were estimated using a Monte Carlo procedure. First, for each spectrum of the data cube, the synthetic stellar spectrum provided by the pPXF technique was subtracted from the original one. Then a histogram of the spectral noise was constructed, based on a spectral region of the residual spectrum without emission lines, and a Gaussian function was fitted to the histogram. After that, Gaussian random noise was created, following a distribution with the same width of the Gaussian fitted to the histogram, and added to the original synthetic stellar spectrum, resulting in ``noisy'' spectra. Finally, the pPXF method was applied to such noisy spectra and the uncertainties of the kinematic parameters were taken as the standard deviations of all the obtained values. 

\begin{figure*}
\begin{center}
   \includegraphics[scale=0.58]{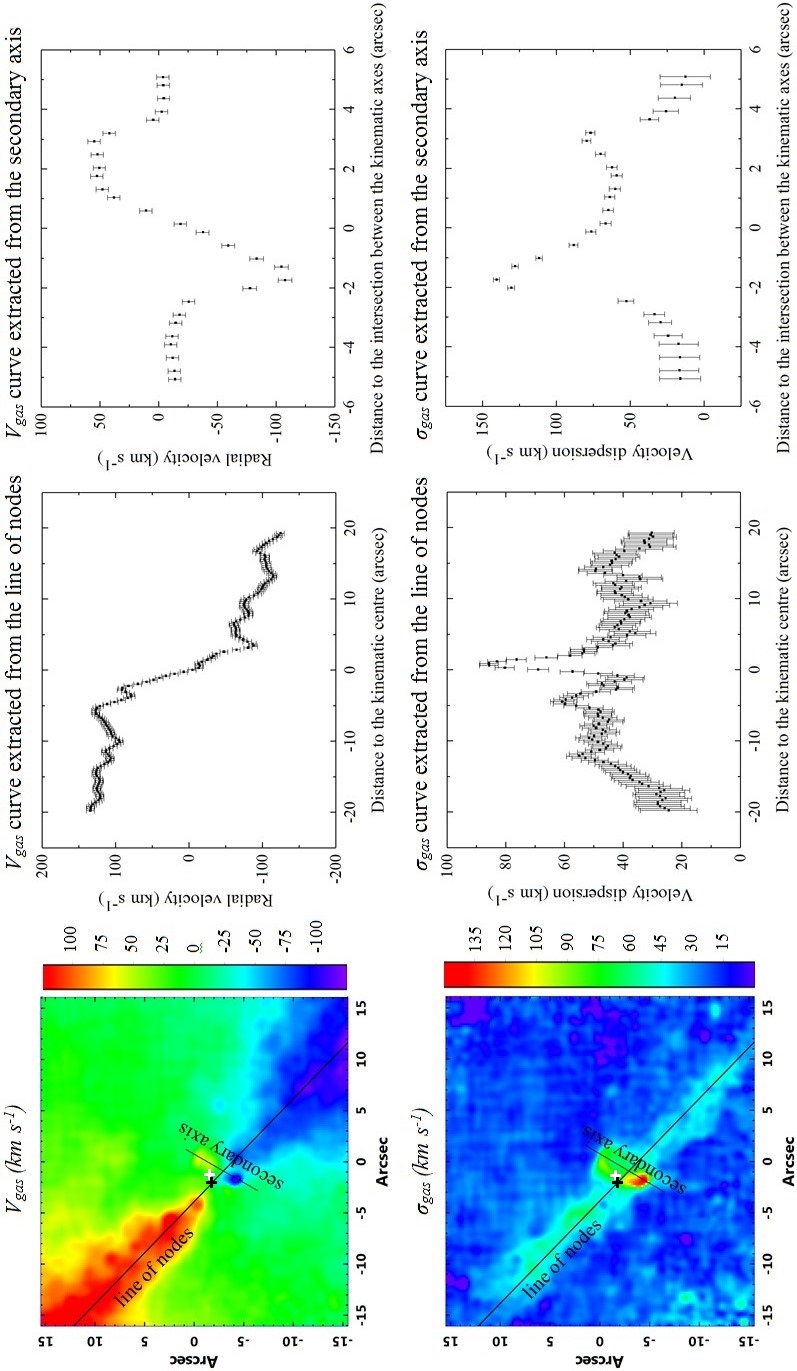} 
  \caption{Maps of the values of the gas radial velocity ($V_{gas}$) and of the gas velocity dispersion ($\sigma_{gas}$), obtained by fitting the [N \textsc{ii}]+H$\alpha$ emission lines in the gas data cube of NGC 1448 with a sum of three Gaussian functions. The curves were extracted from the line of nodes of the $V_{gas}$ map (represented by a black line) and also from the line connecting two compact regions in the $V_{gas}$ map (\textbf{secondary axis,} also represented by a black line) associated with a second kinematic phenomenon. The black cross represents the position of the kinematic centre and the white cross corresponds to the AGN position provided by the hard X-ray emission detected in the Chandra data of this galaxy.}\label{fig12}
\end{center}
\end{figure*}

\section{Gas kinematics}\label{sec6}

The gas kinematic parameters of the central region of NGC 1448 can be obtained by performing a Gaussian fitting of any of the emission lines detected along the FOV of the MUSE data cube, after the subtraction of the stellar continuum. However, considering that the amplitude-to-noise (A/N) ratios of the [N \textsc{ii}]+H$\alpha$ emission lines are considerably high along the entire FOV (and also in order to reduce the uncertainties of the kinematic parameters), we opted to fit the [N \textsc{ii}]+H$\alpha$ emission lines in each spectrum of the data cube with a sum of three Gaussian functions with the same radial velocity ($V_{gas}$) and the same velocity dispersion ($\sigma_{gas}$). Using this procedure, we verified that a minimum A/N ratio of 3 was necessary for the [N \textsc{ii}]$\lambda$6548 emission line to obtain reliable kinematic parameters. In order to reduce the uncertainties of the position angle of the kinematic axis and also of the coordinates of the kinematic centre, we applied this Gaussian fitting procedure to the same MUSE data cube (with a larger FOV) used for the analysis of the stellar kinematics. No spectrum of the data cube with this larger FOV showed a [N \textsc{ii}]$\lambda$6548 emission line with A/N ratio lower than 3. Fig.~\ref{fig12} shows the obtained $V_{gas}$ and $\sigma_{gas}$ maps, together with the $V_{gas}$ and $\sigma_{gas}$ curves extracted along the axis corresponding to the line of nodes of the $V_{gas}$ map. 

A pattern consistent with a gas rotating disc can be easily seen in the $V_{gas}$ map in Fig.~\ref{fig12}. We determined its kinematic centre with the same approach adopted for the $V_*$ map: the point, along the line of nodes, where the $V_{gas}$ value was equal to the average between the maximum and minimum values. This average $V_{gas}$ was subtracted from the $V_{gas}$ map in Fig.~\ref{fig12}, which shows then the velocity values relative to the kinematic centre. The positional uncertainty ($1\sigma$) estimated for this kinematic centre is $0.6$ arcsec. Similarly to the case of the $V_*$ map, the kinematic centre of the $V_{gas}$ map is not coincident with the AGN position, provided by the hard X-ray emission, but such positions are compatible, at the 1$\sigma$ level. The amplitude of the $V_{gas}$ curve is the same of the $V_*$ curve ($\sim 125$ km s$^{-1}$) and the position angle of the axis corresponding to the line of nodes is also the same of the $V_*$ map ($PA_{V_{gas}} = 45\degr \pm 7\degr$). One interesting aspect of the $V_{gas}$ map is that, besides the gas rotation pattern, there appears to be a second kinematic phenomenon, with a compact area, northwest of the AGN, showing gas with redshifted emission and another compact area, south-east of the AGN, showing gas with blueshifted emission. The $V_{gas}$ curve extracted along the axis connecting these two compact regions (secondary axis) is shown in Fig.~\ref{fig12}. The values, in the range of -105 - 55 km s$^{-1}$, are not symmetric relative to the intersection between the line of nodes and the secondary axis. The $\sigma_{gas}$ map shows that the dispersion values in the inner $10$ arcsec of the gas rotating disc are $\sim 50$ km s$^{-1}$. However, in the vicinity of the AGN position, one can note two areas (one northwest of the AGN and the other south-east of the AGN), at the same locations where the compact regions with blueshifted and redshifted emissions were detected, with higher $\sigma_{gas}$ values. The $\sigma_{gas}$ curve extracted along the secondary axis shows that the $\sigma_{gas}$ values in the areas north-west and south-east of the AGN are $\sim 140$ km s$^{-1}$ and $\sim 80$ km s$^{-1}$, respectively. We believe that the second kinematic phenomenon detected in these two compact regions is associated with possible outflows from the AGN. The position angle of the secondary axis is $PA_{outflows} = -32\degr \pm 5\degr$. We did not detect significant asymmetries in the [O \textsc{iii}]$\lambda$5007 emission line, usually associated with outflows. On the other hand, the radial velocity map of the [O \textsc{iii}]$\lambda$5007 emission line revealed essentially the same spatial morphology of the $V_{gas}$ map in Fig.~\ref{fig12}. Considering that, we opted to include in this work only the $V_{gas}$ map obtained by fitting the [N \textsc{ii}]+H$\alpha$ emission lines with a sum of three Gaussian functions. The second gas kinematic phenomenon detected in the maps in Fig.~\ref{fig12} will be discussed in further detail in Section~\ref{sec7}.

The uncertainties of the gas kinematic parameters of each spectrum of the data cube were estimated with a Monte Carlo procedure similar to that used to estimate the uncertainties of the stellar kinematic parameters and consisted of: constructing a histogram of the spectral noise; fitting a Gaussian function to the histogram; generating Gaussian random noise, following a distribution with the same width of the Gaussian fitted to the histogram; generating ``noisy spectra'' corresponding to the sum of the original Gaussians fitted to the emission lines with the random noise; fitting the [N \textsc{ii}]+H$\alpha$ emission lines (with a sum of three Gaussian functions) of the noisy spectra; and calculating the standard deviation of all the obtained kinematic parameters.

\section{Discussion and comparison with previous studies}\label{sec7}

The analysis of the MUSE data cube of the central region of NGC 1448 revealed a variety of physical phenomena. The spatial morphology and classification (based on emission-line ratios) of the line-emitting regions (Figs.~\ref{fig02} and ~\ref{fig05}) are consistent with the presence of two ionization cones around the central AGN (one south-east and the other northwest of the AGN). The emission-line spectrum of the AGN is characteristic of a Seyfert 2. The north-west ionization cone is not so clearly visible, probably due to the high interstellar extinction in the corresponding region. An adequate scenario to be proposed is that these two ionization cones (with an axis with a position angle of $PA_{cones} = -50\degr \pm 7\degr$) are the result of the ionizing radiation emitted by the AGN being collimated by a nearly edge-on obscuring torus, around the BLR, according to the unified model \citep{ant93,urr95}. This is consistent with the findings of \citet{ann17}, who obtained an inclination angle of $78\degr\!\!.6^{+6\degr\!\!.5}_{-11\degr\!\!.8}$ for the obscuring torus by modelling NuSTAR and Chandra X-ray spectra of NGC 1448 with the MYTorus model \citep{mur09}. The high obscuration of the BLR emission caused by a nearly edge-on torus explains (again, according to the unified model) why no broad components of emission lines were detected. This scenario is also consistent with the map showing the classification of the line-emitting regions in Fig.~\ref{fig05}, which reveals that, along a direction nearly perpendicular to the axis of the ionization cones and, therefore, less exposed to the ionizing radiation collimated by the torus, most of the regions show emission-line spectra with a lower ionization degree, typical of LINERs. At larger projected distances of the AGN, most of the areas have emission-line spectra characteristic of transition objects. This type of emission-line spectrum, although somewhat controversial, is usually interpreted as being the result of a LINER-like emission contaminated by the emission from surrounding H II regions \citep{ho93}. In the case of NGC 1448, the transition emission-line spectra may be the result of LINER-like emission from areas farther from the AGN being observed with H II regions along the line of sight. Another possibility is that the gas in such regions is being ionized both by hot young stars and by the ionizing radiation from the AGN. 

Despite the low S/N ratio of the Chandra X-ray data, the nearly spatial correspondence between the soft X-ray emission and Region 2 suggests that the X-ray emitting gas is associated with one of the ionization cones. Previous studies detected extended soft X-ray emission around AGNs (e.g. \citealt{you01,har05,bia06,bal12,gre14,sil20}) and, in certain cases, with a spatial correspondence between the extended X-ray emission and the ionization cone, usually revealed by the [O \textsc{iii}]$\lambda$5007 emission. The most likely hypothesis to explain the X-ray emission assumes that it is originated in the narrow line region (NLR; e.g. \citealt{you01,bia06,bal12,sil20}) and, therefore, it is also the result of the ionization by the central AGN.

The stellar and gas kinematics results indicate a stellar and gas rotating disc around the nucleus, with a position angle of the line of nodes of $PA_{V_*} = PA_{V_{gas}} = 45\degr \pm 7\degr$ and velocity curves showing an amplitude of 125 km s$^{-1}$. The fact that the positions of the kinematic centres obtained for the $V_*$ and $V_{gas}$ maps (Figs.~\ref{fig11} and ~\ref{fig12}) are compatible, at the 1$\sigma$ level, with the AGN position provided by the hard X-ray emission detected in the Chandra data supports the idea that such position corresponds to the actual obscured nucleus of the galaxy. Another result consistent with this scenario is the apparent $\sigma$-drop detected around the AGN position in the $\sigma_*$ map (Fig.~\ref{fig11}). As mentioned in Section~\ref{sec5}, due to the uncertainties of the $\sigma_*$ values, we cannot confirm the existence of this $\sigma$-drop; even so, the featureless continuum emitted by an AGN can result in an obfuscation of the absorption-line spectrum and in lower measured $\sigma_*$ values around the AGN, which can be the case in the central region of NGC 1448. In fact, the map of the flux fractions due to the power law (representing the AGN featureless continuum) provided by the spectral synthesis (Fig.~\ref{fig10}) shows a significant featureless continuum emission in the area of the possible $\sigma$-drop, which is consistent with the proposed scenario. However, such flux fractions are also not totally reliable, due to the degeneracies of the spectral synthesis technique (which, in this case, resulted in featureless continuum being detected in regions farther from the AGN). The results provided by the spectral synthesis without taking into account a power law representing the AGN featureless continuum (Fig.~\ref{fig07}) suggest the presence of a young stellar cluster in the vicinity of the AGN. The presence of such a young nuclear stellar cluster, instead of a featureless continuum emission, could also explain the possible $\sigma$-drop, as these young stars may have formed from kinematically cold gas and may also have kept the low velocity dispersion values. The emission from luminous hot young stars dominates the stellar spectrum, resulting in measured low values for $\sigma_*$ \citep{all05,all06}. Although the $h_3$ values in Fig.~\ref{fig11} are not symmetric relative to the kinematic centre, the apparent anticorrelation between the $h_3$ and $V_*$ maps is usually caused by a stellar rotating disc superposed to a background of stars (the bulge of the galaxy) with radial velocities close to 0.

The most peculiar aspect of the $V_{gas}$ map in Fig.~\ref{fig12} is the presence of a second kinematic phenomenon, with one compact region (north-west of the AGN) showing redshifted emission and another one (south-east of the AGN) showing blueshifted emission. The $\sigma_{gas}$ map in Fig.~\ref{fig12} shows an increase of the values at the positions of the two compact regions, which is probably a consequence of the superposition of the two kinematic phenomena, with distinct radial velocities, along the line of sight resulting in broader emission lines. The position angle of the axis connecting these two compact regions ($PA_{outflows} = -32\degr \pm 5\degr$) is compatible, at the 2$\sigma$ level, with the position angle of the axis of the ionization cones ($PA_{cones} = -50\degr \pm 7\degr$), suggesting that the second kinematic phenomenon is associated with the AGN emission. A natural explanation then is that the second gas kinematic phenomenon corresponds to outflows from the AGN.

The spectral synthesis applied by \citet{ann17} to long-slit spectra of the central region of NGC 1448, considering only stellar populations with solar metallicity, showed that most of the stellar flux ($\sim 75 - 84$\%) is due to stellar populations with 5 Gyr, but young stellar populations (with ages of 5 Myr) were also detected. Such results are relatively similar to the ones provided by the spectral synthesis applied in this work, using a base of stellar population spectra with a larger range of metallicity values, which revealed that $\sim 58.2\%$ of the stellar flux of the MUSE data cube is due to stellar populations with $\sim 1.2 \times 10^{10}$ yr and intermediate metallicity, and $\sim 18.3\%$ of the stellar flux is due to stellar populations with $6.9 \times 10^9$ yr and high metallicity. A young stellar population (with $\sim 3.1 \times 10^6$ yr and low metallicity, associated with $\sim 4.2\%$ of the total stellar flux) was also detected in this work. As mentioned in Section~\ref{sec5}, these spectral synthesis results must be taken with caution, due to the presence of degeneracies and different sources of uncertainties. However, based on the results obtained in this work and also by \citet{ann17}, we can establish that the stellar flux in the central region of NGC 1448 is mostly due to old stellar populations.

The analysis of the Chandra X-ray data indicates that the AGN position (Region 1) is not coincident with the apparent stellar nucleus (Region 3 - Fig.~\ref{fig03}). Actually, the AGN seems to be located at a point with a deficit of optical emission. These results suggest that the position of the true nucleus of the galaxy also corresponds to Region 1. This is consistent with the findings of \citet{ann17}, who called the brightest apparent stellar nucleus the ``optical peak'' and estimated a projected distance of $\sim 3$ arcsec between the AGN and the optical peak. In this work, however, using higher spatial resolution data and based on the AGN position given by the hard X-ray data, we obtain a projected distance of $1.75$ $\pm$ $0.14$ arcsec ($139 \pm 17$ pc) between the AGN and the apparent stellar nucleus. The $A_V$ maps provided by the H$\alpha$/H$\beta$ ratio (Fig.~\ref{fig05}) and by the spectral synthesis (Fig.~\ref{fig09}) indicate high interstellar extinction in the vicinity and northwest of the AGN, probably due to the presence of a dust lane, which causes the obscuration, making the stellar emission from other regions look brighter than that in the region of the true nucleus of the galaxy. \citet{ann17} proposed the same scenario to explain the spatial morphology in the central region of NGC 1448.

One important point to be discussed is the fact that the brightest line-emitting region in the MUSE data cube of NGC 1448 (Region 2) is not located at the AGN position (Region 1). A natural explanation is related to the Compton-thick nature of the AGN, according to the modelling of the NuSTAR and Chandra X-ray spectra performed by \citet{ann17}. In that case, there is a strong obscuration from the AGN towards the observer, but probably not from the AGN towards Region 2. That could explain, in principle, why the line emission is weaker at the AGN position than at Region 2. The $A_V$ values provided by the Balmer decrement of the spectra extracted from regions 1 and 2 are very similar ($A_V(1) = 2.68 \pm 0.18$ mag and $A_V(2) = 2.40 \pm 0.06$ mag), so the higher obscuration towards the AGN position is not due to dust located in the NLR, but due to dust much closer to the AGN, possibly associated with the nearly edge-on torus proposed by \citet{ann17}. On the other hand, a larger covering factor than that of a classical torus may be necessary in order to result in the observed scenario, with little optical emission coming from the AGN position, in comparison to Region 2.

An additional scenario (not mutually exclusive to the one presented above) can be proposed to help to explain the observed emission line properties in the central region of NGC 1448. The deficit of emission at the nucleus may also be a consequence of a recent decrease of the AGN luminosity. In that case, Region 2 is a light ``echo'' or a fossil line-emitting region that is receiving ionizing radiation emitted by the AGN in the past, when it was brighter. \citet{ann17} obtained an intrinsic X-ray luminosity of $L_{2 - 10 keV}$ = ($3.5 - 7.6$) $\times$ $10^{40}$ erg s$^{-1}$ for the AGN in NGC 1448, based on their simulations of NuSTAR and Chandra X-ray spectra. In addition, using the measured intrinsic [O \textsc{iii}] luminosity ($L_{[O \textsc{iii}]}$ = ($6.89 \pm 0.08$) $\times$ $10^{38}$ erg s$^{-1}$) and the X-ray:[O \textsc{iii}] intrinsic luminosity relationship of \citet{pan06}, the authors determined an intrinsic X-ray luminosity of $L_{2 - 10 keV}$ = ($0.2 - 4.5$) $\times$ $10^{40}$ erg s$^{-1}$, which is compatible with the result provided by the modelling of the X-ray spectra, indicating the reliability of the analysis. However, our finding that the AGN position does not coincide with the brightest line-emitting region represents a relevant new information that must be taken into account in this analysis and may be consistent with the fossil line-emitting region hypothesis. Although this work is not focused on the modelling of the emission-line spectrum, we tried to evaluate this new hypothesis by performing a simple simulation with version 13.03 of the \textsc{cloudy} software, last described by \citet{fer13}, in order to try to reproduce the emission-line ratios of the spectrum extracted from the brightest line-emitting region (Region 2). For this simulation, we assumed a central ionizing source emitting a featureless continuum in the form of a power law, with a spectral index of 1.5. We also assumed that the emitting gas is located at a distance from the AGN equal to the observed projected distance between Region 2 and the AGN (135 pc). The electron density of the emitting gas was taken as 215 cm$^{-3}$ (as determined from the [S \textsc{ii}]$\lambda$6717/[S \textsc{ii}]$\lambda$6731 ratio of the spectrum of Region 2, assuming a temperature of 10000 K) and the filling factor was taken as 0.01. We considered a low cut-off energy (taken as a free parameter) for the incident radiation on the emitting gas, to account for the interstellar extinction suffered by the radiation emitted by the AGN, before reaching the line-emitting gas. We repeated the simulation with different values for the bolometric luminosity of the AGN, for the metallicity of the gas and for the cut-off energy of the incident radiation to reproduce the observed emission-line ratios (Table~\ref{tbl2}). At the end, with a bolometric luminosity of $L_{bol} = 1.1 \times 10^{42}$ erg s$^{-1}$, a metallicity of $Z = 0.04$ ($Z_{\sun} = 0.02$ being the solar metallicity) and a cut-off energy of 30 eV, the resulting emission-line ratios were [O \textsc{iii}]$\lambda$5007/H$\beta$ = 8.06 (compatible, at the 1$\sigma$ level, with the value corresponding to Region 2 in Table~\ref{tbl2}), [N \textsc{ii}]$\lambda$6583/H$\alpha$ = 1.15 and [S \textsc{ii}]($\lambda$6717 + $\lambda$6731)/H$\alpha$ = 0.68 (compatible, at the 2$\sigma$ level, with the values corresponding to Region 2 in Table~\ref{tbl2}) and [O \textsc{i}]$\lambda$6300/H$\alpha$ = 0.14 (compatible, at the 3$\sigma$ level, with the value corresponding to Region 2 in Table~\ref{tbl2}). Using the relation $L_{bol} = 16 \times L_{2 - 10 keV}$ \citep{ho08} and the intrinsic X-ray luminosity determined by \citet{ann17} with the modelling of the NuSTAR and Chandra X-ray spectra ($L_{2 - 10 keV}$ = ($3.5 - 7.6$) $\times$ $10^{40}$ erg s$^{-1}$), we obtain a current bolometric luminosity for the AGN in NGC 1448 of $L_{bol}(current)$ = ($5.6 - 12.2$) $\times$ $10^{41}$ erg s$^{-1}$. The upper limit of this luminosity range is consistent with the value provided by our simulation; however the lower limit is equal to 51\% of the value obtained by us. Therefore, if the fossil line-emitting region hypothesis is correct, then the AGN must have reduced its luminosity by nearly half from the epoch when the radiation, now reaching Region 2, was emitted. That was probably the epoch of the origin of the two observed outflows. If we assume that the distance between Region 2 and the AGN is equal to the observed projected distance (135 pc), then any reduction of the AGN luminosity must have occurred during the last $\sim$ 440 yr. It is worth mentioning that, since we detected a possible outflow in the surroundings of Region 2, we evaluated the possibility of shock heating being the mechanism resulting in the observed emission-line spectrum. We checked the Mappings III shock model library (http://cdsweb.u-strasbg.fr/$\sim$allen/mappings\_page1.html - \citealt{all08}). However, considering the obtained $\sigma_{gas}$ values, we concluded that this mechanism is not significant at the central region of NGC 1448. The current bolometric luminosity range of $L_{bol}(current)$ = ($5.6 - 12.2$) $\times$ $10^{41}$ erg s$^{-1}$ is consistent with the luminosities of a significant number of LINERs \citep{ho08}. On the other hand, considering that the nuclear emission-line spectrum of NGC 1448 is characteristic of Seyfert galaxies, we may be observing a transition of AGN types. 

A behaviour similar to the one observed in the central region of NGC 1448 was detected in \citet{men16} in the Gemini Multi-Object Spectrograph/Integral Field Unit data cube of the central region of NGC 3621, which revealed, after a subtraction of the stellar continuum, a deficit of emission at the AGN position and a Seyfert-like compact emission at a projected distance $2.14 \pm 0.08$ arcsec ($70.1 \pm 2.6$ pc) from the AGN. The two hypotheses described above were also proposed to explain these properties. Besides that, such hypotheses were proposed to explain, on a larger scale, the observed emission from the irregular gas cloud SDSS J094103.80+344334.2, known as ``Hanny's Voorwerp'', which is located $\sim$ 25 kpc south-east from the galaxy IC 2497 \citep{joz09,lin09}. The absence of a hard X-ray source at the nucleus of IC 2497, together with the Seyfert-like emission from Hanny's Voorwerp, led to the scenario involving a reduction of the luminosity of the AGN at the nucleus of IC 2497 during the last $10^5$ yr. In this case, Hanny's Voorwerp is still being illuminated by the ionizing radiation emitted by the AGN in the past \citep{lin09,kee12}. The alternative scenario, involving no reduction of the AGN luminosity but instead a high obscuration towards the AGN in IC 2497, was also discussed \citep{joz09,lin09,ram10}. In the specific case of NGC 1448, we believe that an analysis of an X-ray spectrum of the nucleus with a higher S/N ratio would help to support or rule out the fossil line-emitting region hypothesis.

\section{Summary and Conclusions}\label{sec8}

We analysed an optical data cube, obtained with MUSE, of the central region of the late-type galaxy NGC 1448. Our work was focused on the line-emitting regions, on the stellar populations and also on the gas and stellar kinematics around the nucleus of this galaxy. Our main findings are listed below. 

\begin{itemize}

\item The position of the compact hard X-ray source provided by Chandra data indicates that the AGN in NGC 1448 is not located at the apparent stellar nucleus of the galaxy, but at a projected distance of $1.75$ $\pm$ $0.22$ arcsec ($139 \pm 17$ pc) from it. As proposed by previous studies, this is probably a consequence of the high interstellar extinction in the surroundings of the AGN (which corresponds to the true nucleus of the galaxy) that makes the stellar emission from other regions look brighter than that in the true nucleus.

\item The spatial morphology and the classification of the line-emitting regions indicate the presence of a double ionization cone, with an axis with a position angle of $PA_{cones} = -50\degr \pm 7\degr$. This morphology is probably the result of the ionizing radiation from the AGN being collimated by a nearly edge-on torus (identified by previous studies), according to the unified model. The emission-line spectra in the areas inside these ionization cones are compatible with those of Seyfert 2 galaxies. Along a direction perpendicular to the axis of the ionization cones (and, therefore, in areas less exposed to the ionizing radiation from the AGN), the spectra show a lower ionization degree, being characteristic of LINERs.

\item The stellar and gas kinematics are consistent with a stellar and gas rotating disc around the nucleus, with a position angle of the line of nodes of $PA_{V_*} = PA_{V_{gas}} = 45\degr \pm 7\degr$ and with a velocity amplitude of 125 km s$^{-1}$. The kinematic centres of the $V_*$ and $V_{gas}$ maps are compatible, at the 1$\sigma$ level, with the AGN position, which supports the idea that such a position is the actual nucleus of the galaxy.

\item The $\sigma_*$ map reveals an apparent $\sigma$-drop in the surroundings of the AGN. Due to the uncertainties of the $\sigma_*$ values, it is not possible to confirm the presence of such a $\sigma$-drop, which could be the result of an obfuscation of the absorption-line spectrum by the featureless continuum emitted by the AGN or of hot young stars around the AGN that formed from kinematically cold gas and kept the low velocity dispersion values. Both hypotheses are consistent with the results provided by the spectral synthesis. 

\item The gas kinematics results indicate a second kinematic phenomenon (in addition to the gas rotating disc), with one compact region, northwest of the AGN, showing redshifted emission and another compact region, south-east of the AGN, showing blueshifted emission. These two regions show larger $\sigma_{gas}$ values (which is probably a consequence of the superposition of kinematic phenomena with distinct velocities along the line of sight) and the line connecting them has a position angle of $PA_{outflows} = -32\degr \pm 5\degr$. We propose that this kinematic phenomenon is associated with outflows from the AGN.

\item The spectral synthesis results show that most of the stellar flux in the central region of NGC 1448 is due to old stellar populations (with ages of $\sim 1.2 \times 10^{10}$ and $\sim 6.9 \times 10^9$ yr) with intermediate and high metallicity. Lower flux fractions associated with intermediate age ($\sim 6.9 \times 10^9$ yr) and young ($\sim 3.1 \times 10^6$ yr) stellar populations were also detected. This is consistent with results obtained by previous studies.

\item The fact that the AGN position does not coincide with the brightest line-emitting region at the centre of NGC 1448 could be explained by the high obscuration from the AGN towards the observer. The AGN is Compton thick towards the observer, but not towards the brightest line-emitting region. Most of this obscuration is probably caused by a nearly edge-on torus, as determined by previous studies. There is also the additional possibility that the AGN reduced its luminosity and the brightest line-emitting region is actually a fossil or a light echo from the AGN in the past (when it was brighter). Although we cannot confirm this hypothesis, our simulations with the \textsc{cloudy} software indicate that, if the AGN reduced its luminosity, then such a reduction was to 51\% of the luminosity in the past and occurred during the last $\sim 440$ yr. 

\end{itemize}

\section*{Acknowledgements}

This work is based on observations collected at the European Southern Observatory (ESO) under ESO program 097.D-0408. The scientific results reported in this paper are also based in part on data obtained from the Chandra Data Archive under program 15332. This research has made use of the NASA/IPAC Extragalactic Database (NED), which is operated by the Jet Propulsion Laboratory, California Institute of Technology, under contract with the National Aeronautics and Space Administration. We thank Conselho Nacional de Desenvolvimento Cient\'ifico e Tecnol\'ogico (CNPq) for support under grant 306063/2019-0. We also thank an anonymous referee for valuable comments about this paper. 

\section*{Data Availability}

The MUSE data cube analysed in this work is available at the ESO Data Archive (http://archive.eso.org/scienceportal/home). The Chandra data used in this work are available at the Chandra Data Archive (https://cda.harvard.edu/chaser/).



\bibliographystyle{mnras}
\bibliography{menezes} 

\begin{thebibliography}{}
\makeatletter
\relax
\def\mn@urlcharsother{\let\do\@makeother \do\$\do\&\do\#\do\^\do\_\do\%\do\~}
\def\mn@doi{\begingroup\mn@urlcharsother \@ifnextchar [ {\mn@doi@}
  {\mn@doi@[]}}
\def\mn@doi@[#1]#2{\def\@tempa{#1}\ifx\@tempa\@empty \href
  {http://dx.doi.org/#2} {doi:#2}\else \href {http://dx.doi.org/#2} {#1}\fi
  \endgroup}
\def\mn@eprint#1#2{\mn@eprint@#1:#2::\@nil}
\def\mn@eprint@arXiv#1{\href {http://arxiv.org/abs/#1} {{\tt arXiv:#1}}}
\def\mn@eprint@dblp#1{\href {http://dblp.uni-trier.de/rec/bibtex/#1.xml}
  {dblp:#1}}
\def\mn@eprint@#1:#2:#3:#4\@nil{\def\@tempa {#1}\def\@tempb {#2}\def\@tempc
  {#3}\ifx \@tempc \@empty \let \@tempc \@tempb \let \@tempb \@tempa \fi \ifx
  \@tempb \@empty \def\@tempb {arXiv}\fi \@ifundefined
  {mn@eprint@\@tempb}{\@tempb:\@tempc}{\expandafter \expandafter \csname
  mn@eprint@\@tempb\endcsname \expandafter{\@tempc}}}

\bibitem[\protect\citeauthoryear{{Allard}, {Peletier}  \& {Knapen}}{{Allard}
  et~al.}{2005}]{all05}
{Allard} E.~L.,  {Peletier} R.~F.,   {Knapen} J.~H.,  2005, \mn@doi [\apjl]
  {10.1086/498264}, \href
  {https://ui.adsabs.harvard.edu/abs/2005ApJ...633L..25A} {633, L25}

\bibitem[\protect\citeauthoryear{{Allard}, {Knapen}, {Peletier}  \&
  {Sarzi}}{{Allard} et~al.}{2006}]{all06}
{Allard} E.~L.,  {Knapen} J.~H.,  {Peletier} R.~F.,   {Sarzi} M.,  2006,
  \mn@doi [\mnras] {10.1111/j.1365-2966.2006.10751.x}, \href
  {https://ui.adsabs.harvard.edu/abs/2006MNRAS.371.1087A} {371, 1087}

\bibitem[\protect\citeauthoryear{{Allen}, {Groves}, {Dopita}, {Sutherland}  \&
  {Kewley}}{{Allen} et~al.}{2008}]{all08}
{Allen} M.~G.,  {Groves} B.~A.,  {Dopita} M.~A.,  {Sutherland} R.~S.,
  {Kewley} L.~J.,  2008, \mn@doi [\apjs] {10.1086/589652}, \href
  {https://ui.adsabs.harvard.edu/abs/2008ApJS..178...20A} {178, 20}

\bibitem[\protect\citeauthoryear{{Annuar} et~al.,}{{Annuar}
  et~al.}{2017}]{ann17}
{Annuar} A.,  et~al., 2017, \mn@doi [\apj] {10.3847/1538-4357/836/2/165}, \href
  {https://ui.adsabs.harvard.edu/abs/2017ApJ...836..165A} {836, 165}

\bibitem[\protect\citeauthoryear{{Antonucci}}{{Antonucci}}{1993}]{ant93}
{Antonucci} R.,  1993, \mn@doi [\araa] {10.1146/annurev.aa.31.090193.002353},
  \href {https://ui.adsabs.harvard.edu/abs/1993ARA&A..31..473A} {31, 473}

\bibitem[\protect\citeauthoryear{{Bacon} et~al.,}{{Bacon} et~al.}{2010}]{bac10}
{Bacon} R.,  et~al., 2010, in {McLean} I.~S.,  {Ramsay} S.~K.,   {Takami} H.,
  eds,  Society of Photo-Optical Instrumentation Engineers (SPIE) Conference
  Series Vol. 7735, Ground-based and Airborne Instrumentation for Astronomy
  III. p. 773508, \mn@doi{10.1117/12.856027}

\bibitem[\protect\citeauthoryear{{Baldwin}, {Phillips}  \&
  {Terlevich}}{{Baldwin} et~al.}{1981}]{bal81}
{Baldwin} J.~A.,  {Phillips} M.~M.,   {Terlevich} R.,  1981, \mn@doi [\pasp]
  {10.1086/130766}, \href
  {https://ui.adsabs.harvard.edu/abs/1981PASP...93....5B} {93, 5}

\bibitem[\protect\citeauthoryear{{Balmaverde} et~al.,}{{Balmaverde}
  et~al.}{2012}]{bal12}
{Balmaverde} B.,  et~al., 2012, \mn@doi [\aap] {10.1051/0004-6361/201219561},
  \href {https://ui.adsabs.harvard.edu/abs/2012A&A...545A.143B} {545, A143}

\bibitem[\protect\citeauthoryear{{Bianchi}, {Guainazzi}  \&
  {Chiaberge}}{{Bianchi} et~al.}{2006}]{bia06}
{Bianchi} S.,  {Guainazzi} M.,   {Chiaberge} M.,  2006, \mn@doi [\aap]
  {10.1051/0004-6361:20054091}, \href
  {https://ui.adsabs.harvard.edu/abs/2006A&A...448..499B} {448, 499}

\bibitem[\protect\citeauthoryear{{Cappellari} \& {Emsellem}}{{Cappellari} \&
  {Emsellem}}{2004}]{cap04}
{Cappellari} M.,  {Emsellem} E.,  2004, \mn@doi [\pasp] {10.1086/381875}, \href
  {https://ui.adsabs.harvard.edu/abs/2004PASP..116..138C} {116, 138}

\bibitem[\protect\citeauthoryear{{Cardelli}, {Clayton}  \& {Mathis}}{{Cardelli}
  et~al.}{1989}]{car89}
{Cardelli} J.~A.,  {Clayton} G.~C.,   {Mathis} J.~S.,  1989, \mn@doi [\apj]
  {10.1086/167900}, \href
  {https://ui.adsabs.harvard.edu/abs/1989ApJ...345..245C} {345, 245}

\bibitem[\protect\citeauthoryear{{Cid Fernandes}, {Mateus}, {Sodr{\'e}},
  {Stasi{\'n}ska}  \& {Gomes}}{{Cid Fernandes} et~al.}{2005}]{cid05}
{Cid Fernandes} R.,  {Mateus} A.,  {Sodr{\'e}} L.,  {Stasi{\'n}ska} G.,
  {Gomes} J.~M.,  2005, \mn@doi [\mnras] {10.1111/j.1365-2966.2005.08752.x},
  \href {https://ui.adsabs.harvard.edu/abs/2005MNRAS.358..363C} {358, 363}

\bibitem[\protect\citeauthoryear{{Condon}, {Helou}, {Sanders}  \&
  {Soifer}}{{Condon} et~al.}{1996}]{con96}
{Condon} J.~J.,  {Helou} G.,  {Sanders} D.~B.,   {Soifer} B.~T.,  1996, \mn@doi
  [\apjs] {10.1086/192270}, \href
  {https://ui.adsabs.harvard.edu/abs/1996ApJS..103...81C} {103, 81}

\bibitem[\protect\citeauthoryear{{Croton} et~al.,}{{Croton}
  et~al.}{2006}]{cro06}
{Croton} D.~J.,  et~al., 2006, \mn@doi [\mnras]
  {10.1111/j.1365-2966.2005.09675.x}, \href
  {https://ui.adsabs.harvard.edu/abs/2006MNRAS.365...11C} {365, 11}

\bibitem[\protect\citeauthoryear{{Davis} et~al.,}{{Davis} et~al.}{2014}]{dav14}
{Davis} B.~L.,  et~al., 2014, \mn@doi [\apj] {10.1088/0004-637X/789/2/124},
  \href {https://ui.adsabs.harvard.edu/abs/2014ApJ...789..124D} {789, 124}

\bibitem[\protect\citeauthoryear{{Ferland} et~al.,}{{Ferland}
  et~al.}{2013}]{fer13}
{Ferland} G.~J.,  et~al., 2013, \rmxaa, \href
  {https://ui.adsabs.harvard.edu/abs/2013RMxAA..49..137F} {49, 137}

\bibitem[\protect\citeauthoryear{{Ferrarese} \& {Merritt}}{{Ferrarese} \&
  {Merritt}}{2000}]{fer00}
{Ferrarese} L.,  {Merritt} D.,  2000, \mn@doi [\apjl] {10.1086/312838}, \href
  {https://ui.adsabs.harvard.edu/abs/2000ApJ...539L...9F} {539, L9}

\bibitem[\protect\citeauthoryear{{Galbany} et~al.,}{{Galbany}
  et~al.}{2016}]{gal16}
{Galbany} L.,  et~al., 2016, \mn@doi [\mnras] {10.1093/mnras/stv2620}, \href
  {https://ui.adsabs.harvard.edu/abs/2016MNRAS.455.4087G} {455, 4087}

\bibitem[\protect\citeauthoryear{{Gebhardt} et~al.,}{{Gebhardt}
  et~al.}{2000}]{geb00}
{Gebhardt} K.,  et~al., 2000, \mn@doi [\apjl] {10.1086/312840}, \href
  {https://ui.adsabs.harvard.edu/abs/2000ApJ...539L..13G} {539, L13}

\bibitem[\protect\citeauthoryear{{Gonzalez} \& {Woods}}{{Gonzalez} \&
  {Woods}}{2002}]{gon02}
{Gonzalez} R.~C.,  {Woods} R.~E.,  2002, {Digital image processing}

\bibitem[\protect\citeauthoryear{{Goulding} \& {Alexander}}{{Goulding} \&
  {Alexander}}{2009}]{gou09}
{Goulding} A.~D.,  {Alexander} D.~M.,  2009, \mn@doi [\mnras]
  {10.1111/j.1365-2966.2009.15194.x}, \href
  {https://ui.adsabs.harvard.edu/abs/2009MNRAS.398.1165G} {398, 1165}

\bibitem[\protect\citeauthoryear{{Goulding}, {Alexander}, {Lehmer}  \&
  {Mullaney}}{{Goulding} et~al.}{2010}]{gou10}
{Goulding} A.~D.,  {Alexander} D.~M.,  {Lehmer} B.~D.,   {Mullaney} J.~R.,
  2010, \mn@doi [\mnras] {10.1111/j.1365-2966.2010.16700.x}, \href
  {https://ui.adsabs.harvard.edu/abs/2010MNRAS.406..597G} {406, 597}

\bibitem[\protect\citeauthoryear{{Greene}, {Pooley}, {Zakamska}, {Comerford}
  \& {Sun}}{{Greene} et~al.}{2014}]{gre14}
{Greene} J.~E.,  {Pooley} D.,  {Zakamska} N.~L.,  {Comerford} J.~M.,   {Sun}
  A.-L.,  2014, \mn@doi [\apj] {10.1088/0004-637X/788/1/54}, \href
  {https://ui.adsabs.harvard.edu/abs/2014ApJ...788...54G} {788, 54}

\bibitem[\protect\citeauthoryear{{G{\"u}ltekin} et~al.,}{{G{\"u}ltekin}
  et~al.}{2009}]{gul09}
{G{\"u}ltekin} K.,  et~al., 2009, \mn@doi [\apj] {10.1088/0004-637X/698/1/198},
  \href {https://ui.adsabs.harvard.edu/abs/2009ApJ...698..198G} {698, 198}

\bibitem[\protect\citeauthoryear{{Hardcastle} \& {Croston}}{{Hardcastle} \&
  {Croston}}{2005}]{har05}
{Hardcastle} M.~J.,  {Croston} J.~H.,  2005, \mn@doi [\mnras]
  {10.1111/j.1365-2966.2005.09469.x}, \href
  {https://ui.adsabs.harvard.edu/abs/2005MNRAS.363..649H} {363, 649}

\bibitem[\protect\citeauthoryear{{Ho}}{{Ho}}{2008}]{ho08}
{Ho} L.~C.,  2008, \mn@doi [\araa] {10.1146/annurev.astro.45.051806.110546},
  \href {https://ui.adsabs.harvard.edu/abs/2008ARA&A..46..475H} {46, 475}

\bibitem[\protect\citeauthoryear{{Ho}, {Filippenko}  \& {Sargent}}{{Ho}
  et~al.}{1993}]{ho93}
{Ho} L.~C.,  {Filippenko} A.~V.,   {Sargent} W. L.~W.,  1993, \mn@doi [\apj]
  {10.1086/173291}, \href
  {https://ui.adsabs.harvard.edu/abs/1993ApJ...417...63H} {417, 63}

\bibitem[\protect\citeauthoryear{{H{\"o}nig}}{{H{\"o}nig}}{2019}]{hoe19}
{H{\"o}nig} S.~F.,  2019, \mn@doi [\apj] {10.3847/1538-4357/ab4591}, \href
  {https://ui.adsabs.harvard.edu/abs/2019ApJ...884..171H} {884, 171}

\bibitem[\protect\citeauthoryear{{J{\'o}zsa} et~al.,}{{J{\'o}zsa}
  et~al.}{2009}]{joz09}
{J{\'o}zsa} G.~I.~G.,  et~al., 2009, \mn@doi [\aap]
  {10.1051/0004-6361/200912402}, \href
  {https://ui.adsabs.harvard.edu/abs/2009A&A...500L..33J} {500, L33}

\bibitem[\protect\citeauthoryear{{Kauffmann} et~al.,}{{Kauffmann}
  et~al.}{2003}]{kau03}
{Kauffmann} G.,  et~al., 2003, \mn@doi [\mnras]
  {10.1111/j.1365-2966.2003.07154.x}, \href
  {https://ui.adsabs.harvard.edu/abs/2003MNRAS.346.1055K} {346, 1055}

\bibitem[\protect\citeauthoryear{{Keel} et~al.,}{{Keel} et~al.}{2012}]{kee12}
{Keel} W.~C.,  et~al., 2012, \mn@doi [\aj] {10.1088/0004-6256/144/2/66}, \href
  {https://ui.adsabs.harvard.edu/abs/2012AJ....144...66K} {144, 66}

\bibitem[\protect\citeauthoryear{{Kewley}, {Dopita}, {Sutherland}, {Heisler}
  \& {Trevena}}{{Kewley} et~al.}{2001}]{kew01}
{Kewley} L.~J.,  {Dopita} M.~A.,  {Sutherland} R.~S.,  {Heisler} C.~A.,
  {Trevena} J.,  2001, \mn@doi [\apj] {10.1086/321545}, \href
  {https://ui.adsabs.harvard.edu/abs/2001ApJ...556..121K} {556, 121}

\bibitem[\protect\citeauthoryear{{Lintott} et~al.,}{{Lintott}
  et~al.}{2009}]{lin09}
{Lintott} C.~J.,  et~al., 2009, \mn@doi [\mnras]
  {10.1111/j.1365-2966.2009.15299.x}, \href
  {https://ui.adsabs.harvard.edu/abs/2009MNRAS.399..129L} {399, 129}

\bibitem[\protect\citeauthoryear{{Menezes} \& {Steiner}}{{Menezes} \&
  {Steiner}}{2018}]{men18}
{Menezes} R.~B.,  {Steiner} J.~E.,  2018, \mn@doi [\apj]
  {10.3847/1538-4357/aae843}, \href
  {https://ui.adsabs.harvard.edu/abs/2018ApJ...868...67M} {868, 67}

\bibitem[\protect\citeauthoryear{{Menezes}, {Steiner}  \& {Ricci}}{{Menezes}
  et~al.}{2013}]{men13}
{Menezes} R.~B.,  {Steiner} J.~E.,   {Ricci} T.~V.,  2013, \mn@doi [\apjl]
  {10.1088/2041-8205/765/2/L40}, \href
  {https://ui.adsabs.harvard.edu/abs/2013ApJ...765L..40M} {765, L40}

\bibitem[\protect\citeauthoryear{{Menezes}, {Steiner}  \& {Ricci}}{{Menezes}
  et~al.}{2014}]{men14}
{Menezes} R.~B.,  {Steiner} J.~E.,   {Ricci} T.~V.,  2014, \mn@doi [\mnras]
  {10.1093/mnras/stt2381}, \href
  {https://ui.adsabs.harvard.edu/abs/2014MNRAS.438.2597M} {438, 2597}

\bibitem[\protect\citeauthoryear{{Menezes}, {da Silva}, {Ricci}, {Steiner},
  {May}  \& {Borges}}{{Menezes} et~al.}{2015}]{men15}
{Menezes} R.~B.,  {da Silva} P.,  {Ricci} T.~V.,  {Steiner} J.~E.,  {May} D.,
  {Borges} B.~W.,  2015, \mn@doi [\mnras] {10.1093/mnras/stv629}, \href
  {https://ui.adsabs.harvard.edu/abs/2015MNRAS.450..369M} {450, 369}

\bibitem[\protect\citeauthoryear{{Menezes}, {Steiner}  \& {da Silva}}{{Menezes}
  et~al.}{2016}]{men16}
{Menezes} R.~B.,  {Steiner} J.~E.,   {da Silva} P.,  2016, \mn@doi [\apj]
  {10.3847/0004-637X/817/2/150}, \href
  {https://ui.adsabs.harvard.edu/abs/2016ApJ...817..150M} {817, 150}

\bibitem[\protect\citeauthoryear{{Menezes}, {Ricci}, {Steiner}, {da Silva},
  {Ferrari}  \& {Borges}}{{Menezes} et~al.}{2019}]{men19}
{Menezes} R.~B.,  {Ricci} T.~V.,  {Steiner} J.~E.,  {da Silva} P.,  {Ferrari}
  F.,   {Borges} B.~W.,  2019, \mn@doi [\mnras] {10.1093/mnras/sty3337}, \href
  {https://ui.adsabs.harvard.edu/abs/2019MNRAS.483.3700M} {483, 3700}

\bibitem[\protect\citeauthoryear{{M{\"u}ller-S{\'a}nchez},
  {Gonz{\'a}lez-Mart{\'\i}n}, {Fern{\'a}ndez-Ontiveros}, {Acosta-Pulido}  \&
  {Prieto}}{{M{\"u}ller-S{\'a}nchez} et~al.}{2010}]{mul10}
{M{\"u}ller-S{\'a}nchez} F.,  {Gonz{\'a}lez-Mart{\'\i}n} O.,
  {Fern{\'a}ndez-Ontiveros} J.~A.,  {Acosta-Pulido} J.~A.,   {Prieto} M.~A.,
  2010, \mn@doi [\apj] {10.1088/0004-637X/716/2/1166}, \href
  {https://ui.adsabs.harvard.edu/abs/2010ApJ...716.1166M} {716, 1166}

\bibitem[\protect\citeauthoryear{{Murphy} \& {Yaqoob}}{{Murphy} \&
  {Yaqoob}}{2009}]{mur09}
{Murphy} K.~D.,  {Yaqoob} T.,  2009, \mn@doi [\mnras]
  {10.1111/j.1365-2966.2009.15025.x}, \href
  {https://ui.adsabs.harvard.edu/abs/2009MNRAS.397.1549M} {397, 1549}

\bibitem[\protect\citeauthoryear{{Netzer}}{{Netzer}}{2013}]{net13}
{Netzer} H.,  2013, {The Physics and Evolution of Active Galactic Nuclei}

\bibitem[\protect\citeauthoryear{{Osterbrock} \& {Ferland}}{{Osterbrock} \&
  {Ferland}}{2006}]{ost06}
{Osterbrock} D.~E.,  {Ferland} G.~J.,  2006, {Astrophysics of gaseous nebulae
  and active galactic nuclei}

\bibitem[\protect\citeauthoryear{{Panessa}, {Bassani}, {Cappi}, {Dadina},
  {Barcons}, {Carrera}, {Ho}  \& {Iwasawa}}{{Panessa} et~al.}{2006}]{pan06}
{Panessa} F.,  {Bassani} L.,  {Cappi} M.,  {Dadina} M.,  {Barcons} X.,
  {Carrera} F.~J.,  {Ho} L.~C.,   {Iwasawa} K.,  2006, \mn@doi [\aap]
  {10.1051/0004-6361:20064894}, \href
  {https://ui.adsabs.harvard.edu/abs/2006A&A...455..173P} {455, 173}

\bibitem[\protect\citeauthoryear{{Pereira-Santaella}, {Diamond-Stanic},
  {Alonso-Herrero}  \& {Rieke}}{{Pereira-Santaella} et~al.}{2010}]{per10}
{Pereira-Santaella} M.,  {Diamond-Stanic} A.~M.,  {Alonso-Herrero} A.,
  {Rieke} G.~H.,  2010, \mn@doi [\apj] {10.1088/0004-637X/725/2/2270}, \href
  {https://ui.adsabs.harvard.edu/abs/2010ApJ...725.2270P} {725, 2270}

\bibitem[\protect\citeauthoryear{{Prieto}, {Mezcua}, {Fern{\'a}ndez-Ontiveros}
  \& {Schartmann}}{{Prieto} et~al.}{2014}]{pri14}
{Prieto} M.~A.,  {Mezcua} M.,  {Fern{\'a}ndez-Ontiveros} J.~A.,   {Schartmann}
  M.,  2014, \mn@doi [\mnras] {10.1093/mnras/stu1006}, \href
  {https://ui.adsabs.harvard.edu/abs/2014MNRAS.442.2145P} {442, 2145}

\bibitem[\protect\citeauthoryear{{Rampadarath} et~al.,}{{Rampadarath}
  et~al.}{2010}]{ram10}
{Rampadarath} H.,  et~al., 2010, \mn@doi [\aap] {10.1051/0004-6361/201014782},
  \href {https://ui.adsabs.harvard.edu/abs/2010A&A...517L...8R} {517, L8}

\bibitem[\protect\citeauthoryear{{S{\'a}nchez-Bl{\'a}zquez}
  et~al.,}{{S{\'a}nchez-Bl{\'a}zquez} et~al.}{2006}]{san06}
{S{\'a}nchez-Bl{\'a}zquez} P.,  et~al., 2006, \mn@doi [\mnras]
  {10.1111/j.1365-2966.2006.10699.x}, \href
  {https://ui.adsabs.harvard.edu/abs/2006MNRAS.371..703S} {371, 703}

\bibitem[\protect\citeauthoryear{{Schawinski}, {Thomas}, {Sarzi}, {Maraston},
  {Kaviraj}, {Joo}, {Yi}  \& {Silk}}{{Schawinski} et~al.}{2007}]{sch07}
{Schawinski} K.,  {Thomas} D.,  {Sarzi} M.,  {Maraston} C.,  {Kaviraj} S.,
  {Joo} S.-J.,  {Yi} S.~K.,   {Silk} J.,  2007, \mn@doi [\mnras]
  {10.1111/j.1365-2966.2007.12487.x}, \href
  {https://ui.adsabs.harvard.edu/abs/2007MNRAS.382.1415S} {382, 1415}

\bibitem[\protect\citeauthoryear{{Tsai} \& {Hwang}}{{Tsai} \&
  {Hwang}}{2015}]{tsa15}
{Tsai} M.,  {Hwang} C.-Y.,  2015, \mn@doi [\aj] {10.1088/0004-6256/150/2/43},
  \href {https://ui.adsabs.harvard.edu/abs/2015AJ....150...43T} {150, 43}

\bibitem[\protect\citeauthoryear{{Urry} \& {Padovani}}{{Urry} \&
  {Padovani}}{1995}]{urr95}
{Urry} C.~M.,  {Padovani} P.,  1995, \mn@doi [\pasp] {10.1086/133630}, \href
  {https://ui.adsabs.harvard.edu/abs/1995PASP..107..803U} {107, 803}

\bibitem[\protect\citeauthoryear{{Veilleux} \& {Osterbrock}}{{Veilleux} \&
  {Osterbrock}}{1987}]{vei87}
{Veilleux} S.,  {Osterbrock} D.~E.,  1987, \mn@doi [\apjs] {10.1086/191166},
  \href {https://ui.adsabs.harvard.edu/abs/1987ApJS...63..295V} {63, 295}

\bibitem[\protect\citeauthoryear{{Veron-Cetty} \& {Veron}}{{Veron-Cetty} \&
  {Veron}}{1986}]{ver86}
{Veron-Cetty} M.~P.,  {Veron} P.,  1986, \aaps, \href
  {https://ui.adsabs.harvard.edu/abs/1986A&AS...66..335V} {66, 335}

\bibitem[\protect\citeauthoryear{{Young}, {Wilson}  \& {Shopbell}}{{Young}
  et~al.}{2001}]{you01}
{Young} A.~J.,  {Wilson} A.~S.,   {Shopbell} P.~L.,  2001, \mn@doi [\apj]
  {10.1086/321561}, \href
  {https://ui.adsabs.harvard.edu/abs/2001ApJ...556....6Y} {556, 6}

\bibitem[\protect\citeauthoryear{{da Silva}, {Menezes}  \& {Steiner}}{{da
  Silva} et~al.}{2020}]{sil20}
{da Silva} P.,  {Menezes} R.~B.,   {Steiner} J.~E.,  2020, \mn@doi [\mnras]
  {10.1093/mnras/staa007}, \href
  {https://ui.adsabs.harvard.edu/abs/2020MNRAS.492.5121D} {492, 5121}

\bibitem[\protect\citeauthoryear{{de Vaucouleurs}, {de Vaucouleurs}, {Corwin},
  {Buta}, {Paturel}  \& {Fouque}}{{de Vaucouleurs} et~al.}{1991}]{rc3}
{de Vaucouleurs} G.,  {de Vaucouleurs} A.,  {Corwin} Herold~G. J.,  {Buta}
  R.~J.,  {Paturel} G.,   {Fouque} P.,  1991, {Third Reference Catalogue of
  Bright Galaxies}

\makeatother
\end{thebibliography}



\bsp	
\label{lastpage}
\end{document}